\documentclass[twocolumn]{aastex63}
\usepackage{hyperref}
\usepackage{amsmath}
\usepackage{amssymb}
\usepackage{graphicx}
\DeclareRobustCommand{\ion}[2]{\textup{#1\,\textsc{\lowercase{#2}}}}

\newcommand\fei{\ion{Fe}{i}}
\newcommand\feii{\ion{Fe}{ii}}

\newcommand\tii{\ion{Ti}{i}}
\newcommand\tiii{\ion{Ti}{ii}}
\newcommand\oi{\ion{O}{i}}
\newcommand\nai{\ion{Na}{i}}

\newcommand\ali{\ion{Al}{i}}

\newcommand\vi{\ion{V}{i}}
\newcommand\vii{\ion{V}{ii}}
\newcommand\cri{\ion{Cr}{i}}
\newcommand\crii{\ion{Cr}{ii}}

\newcommand{\kms}{km\,s$^{-1}$}
\newcommand{\teff}{$T_{\rm eff}$}
\newcommand{\logg}{$\log g$}
\newcommand{\vt}{$\xi_{t}$}
\newcommand{\micro}{$\nu_{\rm micr}$}
\newcommand{\rproc}{$r$-process}

\newcommand{\AB}[2]{$\mbox{[#1/#2]}$}
\newcommand{\feh}{\AB{Fe}{H}}

\usepackage[utf8]{inputenc}

\received{May 31, 2024}
\revised{July 15, 2024}
\accepted{August 05, 2024}

\submitjournal{ApJS}

\shorttitle{Abundance analysis of faint stars from the RPA}
\shortauthors{Bandyopadhyay et al.}
\begin{document}

\title{The \emph{R}-Process  Alliance: Fifth Data Release from the Search for $R$-Process-Enhanced Metal-poor Stars in the Galactic Halo with the GTC \footnote{This paper includes data gathered with the 10.4 meter Gran Telescopio Canarias located at La Palma, Canary Islands, Spain.}}

\correspondingauthor{Avrajit Bandyopadhyay }
\email{abandyopadhyay@ufl.edu}

\author[0000-0002-8304-5444]{Avrajit Bandyopadhyay}

\affiliation{Department of Astronomy, University of Florida, Bryant Space Science Center, Gainesville, FL 32611, USA}
\affiliation{Joint Institute for Nuclear Astrophysics - Center for Evolution of the Elements, USA}

\author[0000-0002-8504-8470]{Rana Ezzeddine}

\affiliation{Department of Astronomy, University of Florida, Bryant Space Science Center, Gainesville, FL 32611, USA}
\affiliation{Joint Institute for Nuclear Astrophysics - Center for Evolution of the Elements, USA}

\author[0000-0002-0084-572X]{Carlos Allende Prieto}
\affiliation{Instituto de Astrof\'{i}sica de Canarias, V\'{i}a L\'{a}ctea, 38205 La Laguna, Tenerife, Spain}
\affiliation{Universidad de La Laguna, Departamento de Astrofísica, E-38206 La Laguna, Tenerife, Spain}

\author{Nima Aria}
\affiliation{Department of Astronomy, University of Florida, Bryant Space Science Center, Gainesville, FL 32611, USA}
\author[0000-0002-3367-2394]{Shivani P. Shah}
\affiliation{Department of Astronomy, University of Florida, Bryant Space Science Center, Gainesville, FL 32611, USA}

\author[0000-0003-4573-6233]{Timothy C. Beers}
\affiliation{Department of Physics, University of Notre Dame, Notre Dame, IN 46556, USA}
\affiliation{Joint Institute for Nuclear Astrophysics -- Center for Evolution of the Elements, USA}

\author[0000-0002-2139-7145]{Anna Frebel}
\affiliation{Department of Physics \& Kavli Institute for Astrophysics and Space Research, Massachusetts Institute of Technology, Cambridge, MA 02139, USA}
\affiliation{Joint Institute for Nuclear Astrophysics - Center for Evolution of the Elements, USA}

\author[0000-0001-6154-8983]{Terese T. Hansen}
\affiliation{Department of Astronomy, Stockholm University, AlbaNova University Center, SE-106 91 Stockholm, Sweden}
\affiliation{Joint Institute for Nuclear Astrophysics - Center for Evolution of the Elements, USA}

\author[0000-0002-5463-6800]{Erika M. Holmbeck}
\affiliation{Lawrence Livermore National Laboratory, 7000 East Ave, Livermore, CA 94550, USA}
\affiliation{Joint Institute for Nuclear Astrophysics - Center for Evolution of the Elements, USA}

\author[0000-0003-4479-1265]{Vinicius M. Placco}
\affiliation{NSF NOIRLab, Tucson, AZ 85719, USA}
\affiliation{Joint Institute for Nuclear Astrophysics - Center for Evolution of the Elements, USA}

\author[0000-0001-5107-8930]{Ian U. Roederer}
\affiliation{Department of Physics, North Carolina State University; Raleigh, NC, USA}
\affiliation{Joint Institute for Nuclear Astrophysics - Center for Evolution of the Elements, USA}

\author[0000-0002-5095-4000]{Charli M. Sakari}
\affiliation{Department of Physics $\&$ Astronomy, San Francisco State University, San Francisco CA 94132, USA}
\affiliation{Joint Institute for Nuclear Astrophysics - Center for Evolution of the Elements, USA}

\begin{abstract}

Understanding the abundance pattern of metal-poor stars and the production of heavy elements through various nucleosynthesis processes offers crucial insights into the chemical evolution of the Milky Way, revealing primary sites and major sources of rapid neutron-capture process ($r$-process) material in the Universe. In this fifth data release from the \emph{R}-Process Alliance, we present the detailed chemical abundances of 41 faint (down to $V = 15.8$) and extremely metal-poor (down to [Fe/H] = $-3.3$) halo stars selected from the $R$-Process Alliance (RPA). We  obtained high-resolution spectra for these objects with the HORuS spectrograph on the Gran Telescopio Canarias. We measure the abundances of light, $\alpha$, Fe-peak, and neutron-capture elements. We report the discovery of five CEMP, one limited-$r$, three $r$-I, and four $r$-II stars, and six Mg-poor stars. We also identify one star of a possible globular cluster origin at an extremely low metallicity at [Fe/H] $= -3.0$. This adds to the growing evidence of a lower limit metallicity floor for globular cluster abundances. We use the abundances of Fe-peak elements and the $\alpha$-elements to investigate the contributions from different nucleosynthesis channels in the progenitor supernovae. We find the distribution of [Mg/Eu] as a function of [Fe/H] to have different enrichment levels, indicating different possible pathways and sites of their production. We also reveal differences in the trends of the neutron-capture element abundances of Sr, Ba, and Eu of various $r$-I and $r$-II stars from the RPA data releases, which provide constraints on their nucleosynthesis sites and subsequent evolution.

\end{abstract}




\section{Introduction}\label{sec:int}

Following the Big Bang, the cosmic primordial gas was  composed of H and He, with traces of Li. The first stars that lit up the Universe were free of metals, but after exploding as 
supernovae, they introduced newly synthesized metals to their local interstellar medium (ISM) \citep{christlieb2002,Beers2005,caffau2013,frebel2013,spite2013,bonifacio2015,roederer2016,Vanni2024}. This resulted in significant impacts on not only the evolution of their local ISM, but also affected mini-halos located relatively far from their explosion sites \citep{caffau2011,cooke2014,rodererjacobson2014}. Recurring supernovae events led to the gradual enrichment of the ISM with time; subsequent generations of stars were formed from gas clouds that included heavy elements from the previous stellar generations.

Low-mass, metal-poor stars are among the oldest stellar populations, and are still observable today in the halo of the Milky Way (MW) 
\citep{beers1985,beers1992,barklem2002,christlieb2002,Beers2005,cohen2013,Frebel2015}. These stars provide a unique opportunity to look back in time to study the nucleosynthesis events that took place in the early Galaxy. The atmospheres of very metal-poor (VMP; [Fe/H] $\leq -2.0$) and extremely metal-poor (EMP; [Fe/H] $\leq -3.0$)\footnote{[A/B] = $\log(N_A/{}N_B)_{\star} - \log(N_A/{}N_B) _{\odot}$, where $N$ is the number density of atoms of a given element in the star ($\star$) and the Sun ($\odot$), respectively.} stars retain the abundance signatures of Population III stars and the imprints of the nucleosynthesis processes that occurred during the explosions and in stellar winds \citep{Beers2005,mcwilliam2018}. Although these low-mass, metal-poor stars trace chemical evolution from the earliest times, their observed abundance patterns reflect contributions from multiple stellar generations rather than exclusively from the first stellar generation's yields and initial mass function (IMF). The relative abundances of the elements measured in these stars, which formed in different sites after different nucleosynthesis processes had enriched the birth gas clouds, hold the keys to deciphering the physical events that occurred in the early MW \citep{freb2014r,kobayashi2020,Lunney2020,Arcones2023}. However atomic diffusion and non-canonical stellar processes have minimal impact on low-metallicity stars due to their shallower outer convection zones and reduced efficiency \citep{Spite2005,Korn2007,Lind2008} and hence we assume that the low-mass ancient stars we observe today retain Population III abundance signatures, as supported by consistent spectroscopic observations and stellar evolution models.

The various elemental-production sites contribute to different regions of the periodic table, and are often unrelated to each other \citep{chiaki2012,johnson2019,Roe2022}. At the earliest epochs, the odd-Z elements are produced in massive stars as well as core-collapse supernovae (CCSNe); the $\alpha$- and Fe-peak elements are produced in several sites, such as hydrostatic and explosive burning phases of CCSNe, hypernovae (HNe), and pair instability supernovae (PISNe) \citep{nakamura1999,heger2002,heger2010,nomoto2013}. The predicted relative yields of the different elements produced by these classes of progenitors differ from one another in a number of ways. For instance, a strong “odd/even” effect \citep{heger2002} is expected to be found, along with low [Zn/Fe] ratios, in the ejecta of very massive objects exploding as PISNe, which is less-pronounced for the case of CCSNe \citep{cayrel2001,cayrel2004,cohen2004}. Measuring accurate estimates from VMP/EMP stars hold the key to understanding and disentangling the nature of possible contributors to their overall enrichment.

The production of the elements beyond the Fe peak primarily occurs via three routes -- the slow ($s$-), intermediate ($i$-), and rapid ($r$-) neutron-capture processes. While the origin for the $s$-process in AGB stars is relatively well-understood  \citep{gallino1998,Busso2001,Karlug2016,frebel-rev2018}, a number of sites for the \rproc, such as binary neutron star mergers (NSMs) \citep{lattimer1974}, magneto-rotationally driven jets \citep{winteler2012}, or collapsar disk winds \citep{siegel2019,Braue_2021}  have been proposed over the last few decades, but no consensus has been reached thus far \citep{cote2019}. However, NSMs are the only sites to have observational evidence for hosting the $r$-process so far. The $i$-process \citep{Cowan77,Hampel2016} is more commonly associated with the early AGB phase of low-metallicity, low-mass stars, resulting from the ingestion of protons in a convective helium-burning region \citep{Choplin2021,choplin22}. Furthermore, the $rp$ process, particularly in conjunction with photodisintegration, contributes to the synthesis of elements beyond Fe, elucidating the production mechanism of rare proton-rich isotopes such as $^{92}$Mo \citep{Arcones23}.

The $R$-Process Alliance (RPA) collaboration aims to significantly increase the number of observed \rproc-enhanced (RPE) stars through the detailed study of neutron-capture elements, along with light, $\alpha$-, and Fe-peak elements. This comprehensive approach seeks to understand the formation sites of these stars and the processes that enriched their birth gas clouds. Additionally, this research will provide insights and constraints on the production of these different groups of elements. The primary motivation of the RPA is to combine observations, theoretical advances, and results from chemical-evolution simulations to eventually produce a more complete understanding of the origin of the RPE stellar population in the MW. To this end, four data releases have  been published (RPA-1: \cite{rpa1}, RPA-2: \cite{rpa2}, RPA-3: \cite{rpa3}, RPA-4:  \cite{rpa4}); these papers report dozens of newly discovered RPE stars. The present study is the fifth data release from the RPA (RPA-5) and has targeted the fainter stars selected from the RPA sample for follow-up spectroscopy with the 10.4\,m Gran Telescopio Canarias. Along with the key neutron-capture elements Sr, Ba, and Eu, these papers utilize the abundances of all the observed elements from C to Zn to understand the chemical evolution of the MW.

This paper is outlined as follows.  Section\,\ref{sec:obs} describes the observations, data reduction, and radial-velocity measurements. Section\,\ref{sec:stell_param} presents determinations of the stellar parameters of the sample using 1D (1-Dimensional), LTE (Local Thermodynamic Equilibrium) stellar-atmosphere models, and the necessary corrections to the adopted values. Section\,\ref{sec:abund} presents the chemical abundances of detected light, $\alpha$-, Fe-peak, and neutron-capture elements for the sample stars. Finally, we  discuss our results and conclusions in Sections\,\ref{sec:disc} and \ref{sec:conc}, respectively.

\section{Observations}\label{sec:obs}

\subsection{Target Selection and Observations}

The observing program was carried out as a part of the RPA ``snapshot" survey efforts, during which moderately high-resolution ($R \sim 30,000$) spectra at intermediate signal-to-noise ratios (SNR; $\sim 30$) are obtained in order to identify new RPE stars. The target stars had been selected from various low- ($R \sim 1800$) and medium- ($R \sim 7500$) resolution spectroscopic surveys for metal-poor stars in the Galaxy, including the Large Sky Area Multi-Object Fibre Spectroscopic Telescope (LAMOST; \citealp{2012lamost}) and the RAdial Velocity Experiment (RAVE; \citealp{RAVE2006}) surveys, among others. The metal-poor nature had been determined by \citet{placco2018} based on these medium-resolution spectra. Additional details on the selection criteria are provided in \citet{placco2018,placco2019} and \citet{rpa1,rpa3}.

\begin{figure}
\centering
\hspace*{-1cm}
\includegraphics[scale=0.4]{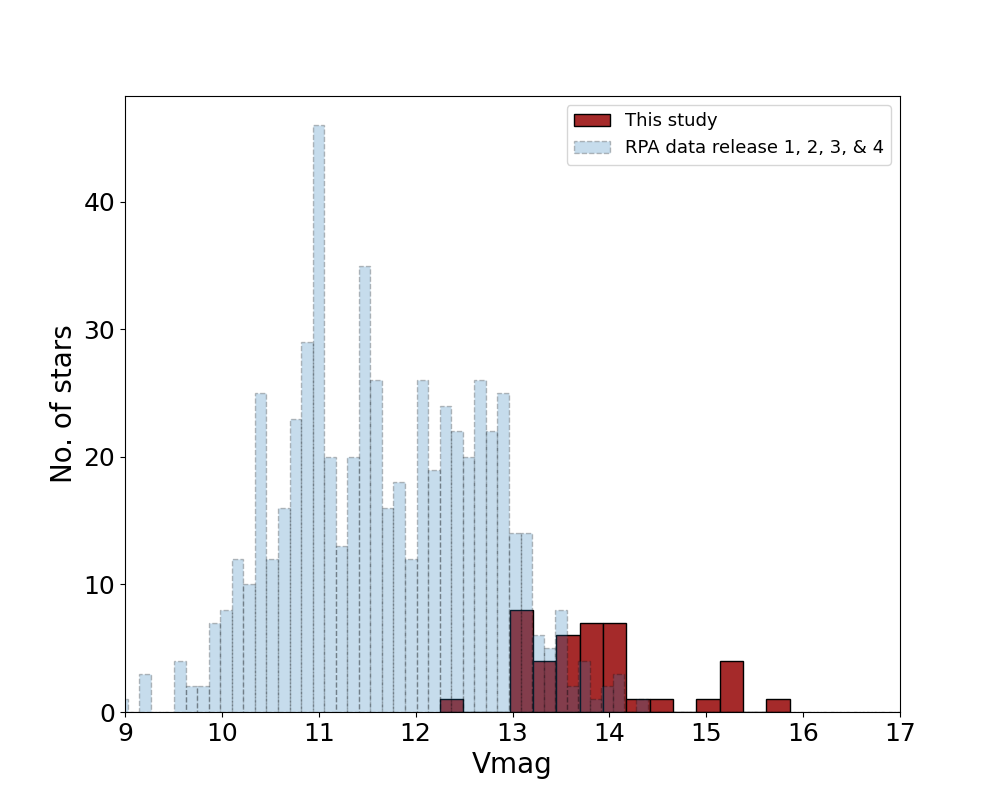}
\caption{Distribution of the $V$ magnitudes for the current sample of stars are shown in the red histogram. The $V$ magnitudes of the stars in this study lie between 13.0 and 15.8, making it the faintest RPA sample of stars by more than one magnitude. The cumulative RPA samples \citep{rpa1,rpa2,rpa3,rpa4} are shown in the background (light-blue histogram).}
\label{Vmag_hist}
\end{figure}

These targets were then observed at a spectral resolving power of $R \sim 25,000$, using the High Optical Resolution Spectrograph (HORuS) \citep{carlos2020} on the  10.4 meter Gran Telescopio Canarias (GTC) located on La Palma, Canary Islands, Spain. Due to the large aperture of the GTC and favorable observing conditions, relatively fainter targets could be studied. In comparison to the previous RPA samples \citep{rpa1,rpa2,rpa3,rpa4} that studied stars with $V < 14.2$, the current stars extend to $V = 15.8$, as shown in Figure \ref{Vmag_hist}. 
Spectra for 45 metal-poor stars were obtained as a part of the FILLER program on the GTC in 2020. Data for four objects had to be discarded due to poor quality, reducing the number of stars in this study to 41. The exposure times varied between 600s and 2000s, depending on the stellar magnitude and weather conditions. The SNR ranges between 9 and 40 with a mean SNR of 26 at 5000\,{\AA}. The low SNR in the blue region does not allow us to calculate the precise abundances for a large number of neutron-capture elements, but those for the key elements Sr, Ba, and Eu could still be derived (or have meaningful upper limits determined). 
The Two Micron All Sky Survey (2MASS) IDs, Right Ascension (R.A.) and declination (Dec), visual magnitudes ($V$), near-infrared magnitudes ($J$), exposure times, SNR at 5000\,{\AA}, and radial velocities from $Gaia$ (RV$_{\rm Gaia}$) and our spectra (RV$_{\rm helio}$) are listed in Table \ref{tab:ident}.

\begin{table*}
    \centering
    \caption{Observational Details of the Target Stars}
    \label{tab:ident}
    
    \begin{tabular}{lccccccrr} 
        \hline
        Name & R.A. &Dec & $V$ mag & $J$ mag & Exp. time & SNR & \multicolumn{1}{r}{RV$_\mathrm{Gaia}$} & \multicolumn{1}{r}{RV$_\mathrm{helio}$}\\ 
        & &  & & & (sec) & &(km s$^{-1}$) &(km s$^{-1}$)\\
		\hline
2MASS J00125284+4726278 &00:12:52.848 &+47:26:27.84 &13.85 &11.72 &1600 &29 &$-$80.1 &$-$80.3\\ 
2MASS J01171437+2911580 &01:17:14.371 &+29:11:57.98 &13.52 &11.61 &900 &38 &$-$136.3 &$-$137.9\\  
2MASS J01261714+2620558 &01:26:17.139 &+26:20:55.84 &13.96 &11.97 &1200 &11 &$-$162.4 &$-$171.4\\  
2MASS J02462013$-$1518418\tablenotemark{a} &02:46:20.130 &$-$15:18:41.80 &10.70 &12.30 &1400 &34 &278.5 &259.3\\
2MASS J04051243+2141326 &04:05:12.430 &+21:41:32.64 &13.51 &11.53 &1100 &34 &\dots &$-$310.9\\ 
2MASS J04464970+2124561 &04:46:49.709 &+21:24:56.02 &15.22 &13.21 &2000 &19 &$-$59.1 &$-$54.2\\  
2MASS J05455436+4420133\tablenotemark{b} &05:45:54.367 &+44:20:13.34 &13.42 &10.88 &1200 &18 &$-$67.1 &$-$87.9\\
2MASS J06114434+1151292\tablenotemark{a} &06:11:44.340 &+11:51:29.20 &\dots &10.40 &900 &14 &341.3 &305.2\\ 
2MASS J06321853+3547202 &06:32:18.530 &+35:47:20.20 &13.78 &11.88 &900 &37 &$-$81.9 &$-$85.5\\  
2MASS J07424682+3533180 &07:42:46.822 &+35:33:17.92 &13.96 &12.22 &1200 &23 &216.3 &216.2\\  
2MASS J07532819+2350207 &07:53:28.198 &+23:50:20.66&13.78 &12.00 &900 &29 &353.4 &353.4\\   
2MASS J08011752+4530033 &08:01:17.505 &+45:30:03.42 &13.26 &11.57 &900 &48 &47.2 &51.2\\  
2MASS J08203890+3619470 &08:20:38.911 &+36:19:47.02 &15.81 &13.64 &1800 &13 &\dots &$-$64.5\\  
2MASS J08471988+3209297 &08:47:19.885 &+32:09:29.77 &13.70 &11.37 &900 &14 &$-$21.7 &$-$22.5\\  
2MASS J09092839+1704521 &09:09:28.395 &+17:04:52.17 &14.88 &13.42 &1800 &18 &123.1 &123.8\\  
2MASS J09143307+2351544 &09:14:33.076 &+23:51:54.40 &13.22 &11.26 &900 &39 &$-$48.3 &$-$46.1\\  
2MASS J09185208+5107215 &09:18:52.082 &+51:07:21.37&13.09 &11.45 &600 &40 &$-$53.4 &$-$48.2\\   
2MASS J09261148+1802142 &09:26:11.477 &+18:02:14.44&14.60 &12.49 &1500 &29 &199.5 &198.9\\   
2MASS J09563630+5953170 &09:56:36.309 &+59:53:17.06 &13.36 &10.99 &900 &20 &$-$285.9 &$-$285.6\\  
2MASS J10122279+2716094 &10:12:22.792 &+27:16:09.43&15.19 &13.52 &1800 &18 &\dots &34.2\\   
2MASS J10542923+2056561 &10:54:29.231 &+20:56:55.91 &14.20 &12.41 &1200 &09 &88.3 &89.4\\  
2MASS J11052721+3305150 &11:05:27.221 &+33:05:15.08&13.77 &12.18 &900 &34 &$-$205.2 &$-$205.9\\   
2MASS J12131230+2506598 &12:13:12.305 &+25:06:59.87 &13.82 &12.17 &900 &22 &$-$91.1 &$-$88.8\\  
2MASS J12334194+1952177 &12:33:41.935 &+19:52:17.59 &13.00 &10.89 &600 &37 &68.5 &68.7\\  
2MASS J12445815+5820391 &12:44:58.178 &+58:20:39.13 &13.76 &11.72 &900 &18 &$-$67.5 &$-$67.9\\  
2MASS J13281307+5503080 &13:28:13.077 &+55:03:07.99 &13.48 &12.34 &800 &11 &$-$0.5 &2.9\\  
2MASS J13525684+2243314 &13:52:56.851 &+22:43:31.55 &13.59 &11.31 &900 &16 &5.1 &6.7\\  
2MASS J13545109+3820077 &13:54:51.097 &+38:20:07.81&13.74 &11.88 &900 &22 &141.4 &130.2\\   
2MASS J14245543+2707241 &14:24:55.435 &+27:07:24.18&15.29 &13.74 &2000 &17 &\dots &19.7\\   
2MASS J14445238+4038527 &14:44:52.377 &+40:38:52.72 &13.03 &10.93 &600 &12 &$-$104.5 &$-$112.3\\  
2MASS J15442141+5735135\tablenotemark{b}&15:44:21.414 &+57:35:13.51 &13.97 &12.10 &1200 &17 &$-$143.5 &$-$129.6\\  
2MASS J16374570+3230413 &16:37:45.696 &+32:30:41.20 &13.41 &11.83 &900 &22 &$-$234.4 &$-$227.2\\  
2MASS J16380702+4059136 &16:38:07.029 &+40:59:13.68&14.01 &12.56 &1800 &26 &$-$25.3 &$-$15.4\\   
2MASS J16393877+3616077 &16:39:38.767 &+36:16:07.66 &13.13 &11.58 &900 &24 &$-$116.2 &$-$108.2\\  
2MASS J16451495+4357120 &16:45:14.952 &+43:57:12.05 &13.10 &11.26 &600 &19 &$-$84.1 &$-$77.0\\  
2MASS J17041197+1626552 &17:04:11.974 &+16:26:55.20&13.93 &11.86 &900 &18 &$-$176.6 &$-$171.0\\   
2MASS J17045729+3720576 &17:04:57.300 &+37:20:57.62 &14.12 &12.02 &1200 &19 &$-$152.1 &$-$148.0\\  
2MASS J17125701+4432051 &17:12:57.021 &+44:32:05.16 &13.42 &11.34 &900 &24 &$-$123.7 &$-$121.2\\  
2MASS J21463220+2456393 &21:46:32.210 &+24:56:39.42&15.29 &13.46 &1800 &12 &\dots &$-$309.3\\   
2MASS J22175058+2104371 &22:17:50.588 &+21:04:37.19&13.39 &11.30 &1200 &29 &\dots &$-$114.6\\   
2MASS J22424551+2720245 &22:42:45.505 &+27:20:24.54&13.14 &11.29 &1800 &40 &\dots &$-$392.2 \\ 
	\hline
	\end{tabular}
\tablenotetext{a}{Indicates likely binary star, based on the reported RUWE from $Gaia$.}
\tablenotetext{b}{Indicates potential binary star, based on deviations in RV of more than 1$\sigma$ between the $Gaia$ RVs and our determination.}
\end{table*}

\subsection{Data Reduction and Radial Velocities}

\begin{figure}
\centering
\hspace*{-0.5cm}
\includegraphics[scale=0.37]{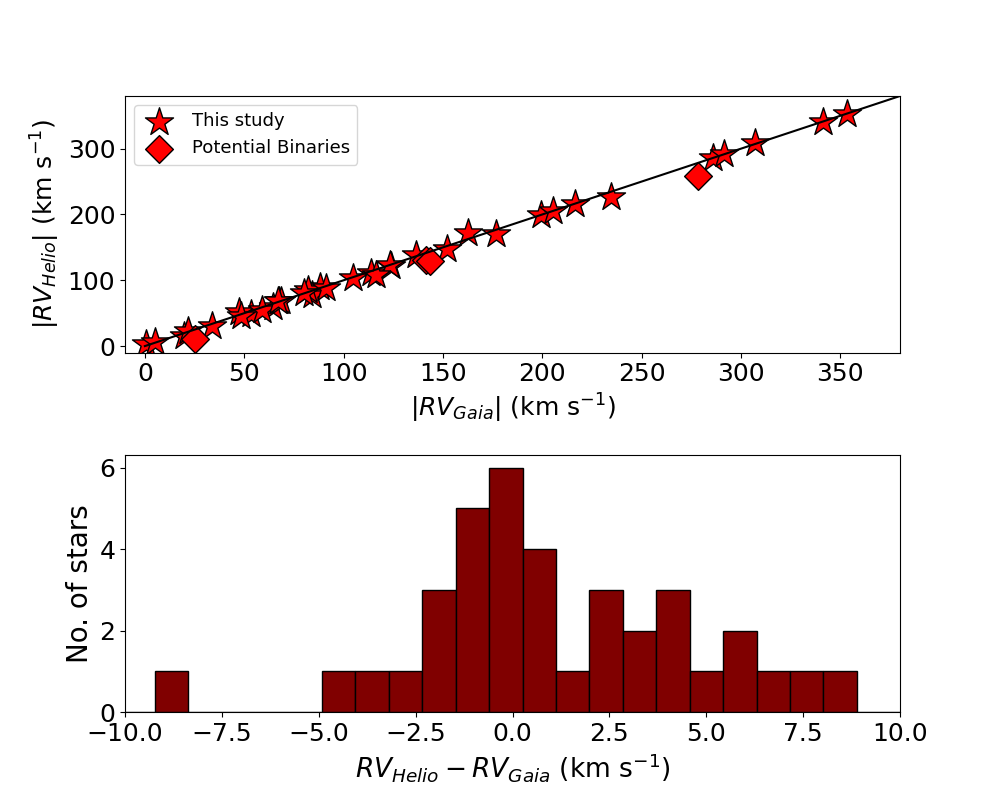}
\caption{Top panel:  Heliocentric radial velocities for our sample stars compared to the $Gaia$ RVs.  The black line is the one-to-one line. The diamond symbols indicate potential binaries. Lower panel: Differences between the two quantities.} 

\label{drv}
\end{figure}

The spectra were reduced using the dedicated HORuS pipeline \texttt{chain}\footnote{https://github.com/callendeprieto/chain},\footnote{https://github.com/callendeprieto/chain/releases/tag/RPA2024}, which includes sky subtraction, tracing of individual orders, wavelength calibration, and continuum normalization. The individual extracted and normalized orders were then merged to produce a final spectrum for each star. The final spectra were analyzed using the \texttt{Spectroscopy Made Harder} (\texttt{SMHr}) software (first described in \citealt{casey2014}). The radial velocities were determined via SMHr using cross-correlation of prominent lines throughout the spectra with those of well-studied standard stars of similar evolutionary stages. 
Heliocentric radial velocities (RV$_{\mathrm{helio}}$) were then determined with the \texttt{rvcorrect} package in \texttt{PYRAF}. The final derived values are listed in Table\,\ref{tab:ident}. For the majority of the stars with available $Gaia$ RVs, the values agree well (mean deviation of 2 km s$^{-1}$ and standard deviation of 5 km s$^{-1}$). An RV comparison is shown in Figure \ref{drv}. In the top panel, heliocentric velocities are compared to the $Gaia$ RVs. Differences between the two measurements is shown as a histogram in the bottom panel. We note that two of the stars (2MASS J06114434+1151292 and 2MASS J02462013-1518418) with renormalised unit weight error (RUWE) of 1.2 from $Gaia$ are expected to be binaries; they exhibit large deviations in RV, on the order of 15 km s$^{-1}$. These two objects are not considered for calculating the mean and standard deviations for the RVs mentioned above, but are included in rest of the paper. There are two additional candidates for binarity found in this study, with RV variations larger than 1$\sigma$.

\begin{figure}
\centering
\hspace*{-0.6cm}
\includegraphics[scale=0.4]{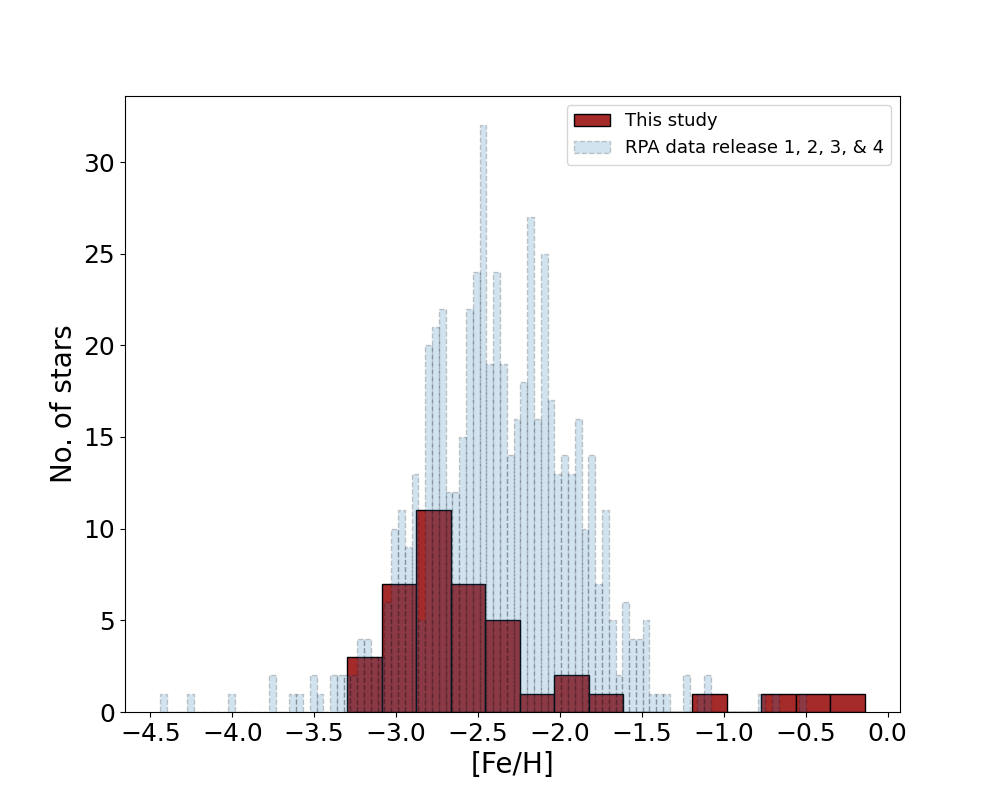}
\caption{Metallicity distribution of our sample of stars, shown as the red histogram. Stars from all other previous RPA data releases are shown in the background (light-blue) histogram. The current sample spans a metallicity range of $-3.3 \leq$ [Fe/H] $ \leq -0.2$, with a peak at [Fe/H] $=- 2.8$. }
\label{feh_hist}
\end{figure}

\section{Stellar Atmospheric Parameters}\label{sec:stell_param}

Stellar atmospheric parameters for our sample stars (effective temperature, \teff; surface gravity, \logg; 
metallicity, \feh; microturbulent velocity, \vt) were derived from measurements of equivalent widths (EW) of \fei\ and \feii\ lines. The equivalent widths of the Fe lines were measured by fitting Gaussian line profiles to the spectral absorption features using \texttt{SMHr}. The initial LTE stellar atmospheric parameters were estimated from the abundances of $\fei$ and $\feii$ lines, using the LTE radiative transfer code \texttt{MOOG} \citep{sneden1973}, including Rayleigh scattering treatment (following \citealt{sobeck2011}\footnote{https://github.com/alexji/moog17scat}).  The 1D, LTE stellar atmospheric ATLAS models employed are from \citet{castelli2004}, with a standard $\alpha$-element enhancement of [$\alpha$/Fe] = +0.4. 

Initial estimates for \teff\ were derived following the principle of excitation equilibrium, by demanding that there be no trend of $\fei$ line abundances with excitation potential. We also enforced the principle of ionization equilibrium by varying \logg\ until we obtain the same abundances from both $\fei$ and $\feii$ lines. The \vt\ was determined by ensuring that there be no trends for $\fei$ abundances with reduced equivalent widths. The \feh\ values were determined from the  mean of $\fei$ and $\feii$ lines, after estimation of the LTE parameters for \teff\,, \logg\,, and \vt. 

We next revised the LTE spectroscopic stellar parameters, as they are known to result in cooler temperatures and low surface-gravity estimates, due to several reasons (imperfect treatment of scattering, impact of approximations when modeling the line formation, wavelength coverage,
data quality, and non-LTE effects), as discussed in \cite{frebel2013}. The corrected \teff\ were determined following the empirical calibration of the derived \teff\ to a photometric scale as given by \cite{frebel2013}:

\begin{displaymath}
    {T_{\mathrm{eff}}({\rm FR13}_{\rm corr}) = {0.9\times} {T_{\mathrm{eff}}\text{(LTE)}} + 670}
\end{displaymath}

After deriving the corrected \teff, denoted as \\\teff(FR13$_{\rm corr}$), we re-derived \logg, \vt, and [Fe/H]. The FR13 correction resulted in warmer temperatures with higher \logg\ and [Fe/H] for the target stars. Estimates for the stellar parameters using both LTE and FR13$_{\rm corr}$ parameters are listed in Table \ref{tab:stell_param}.  As discussed in 
\citet{frebel2013}, \citet{ezzeddine2017}, and \citet{rpa3}, the FR13$_{\rm corr}$ parameters are more reliable approximations of the stellar parameters. Hence, going forward in this study, we adopt the 
FR13$_{\rm corr}$ stellar parameters to derive the abundances. 

\begin{figure}
\centering
\hspace*{-0.8cm}
\includegraphics[scale=0.4]{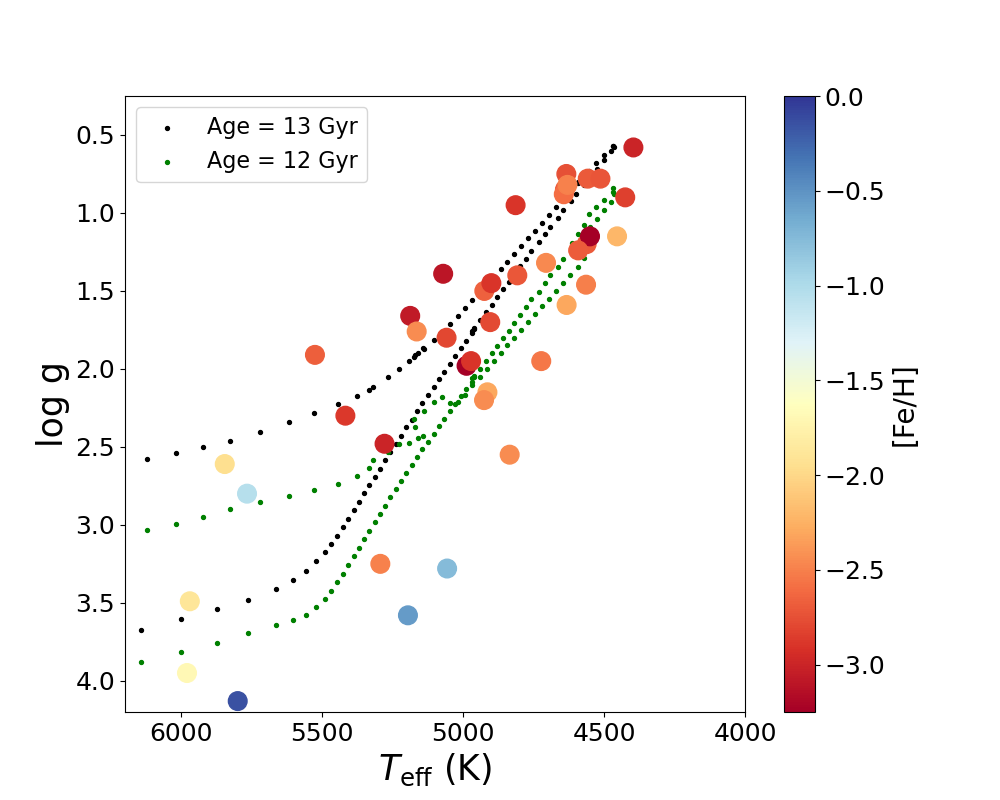}
\caption{The HR diagram showing the sample stars color-coded by metallicity, as indicated by the color bar. The stellar evolutionary tracks correspond to ages of 12 (blue-dotted lines) and 13 Gyr 
(green-dotted lines) for a metallicity of [Fe/H] $= -2.3$.}
\label{kiel}
\end{figure}

The FR13$_{\rm corr}$ metallicity distribution of our sample stars is shown in Figure \ref{feh_hist}. The metallicity ranges from [Fe/H] = $-$3.2 to [Fe/H] = $-0.2$, with a peak at [Fe/H] = $-$2.8. The metallicity distribution largely covers  the same range of the previously published RPA data releases, but it peaks at a slightly lower metallicity. Figure \ref{kiel} shows the distribution of \teff\ and \logg, color-coded by \feh, for our sample stars. The data are overlaid by isochrones \footnote{http://stev.oapd.inaf.it/cgi-bin/cmd} \citep{Marigo2017} for [Fe/H] = $-$2.4 and ages corresponding to 12 and 13 Gyr.  From inspection, the current sample includes stars on the main sequence, main-sequence turnoff, subgiant branch, and approaching the tip of the red giant branch.

\startlongtable
\begin{deluxetable*}{l c c c c c c c c c c c c}
\tablewidth{1000pt}
\tabletypesize{\footnotesize}
\tablecaption {Stellar Atmospheric Parameters of the Target Stars \label{tab:stell_param}} 
\tablehead{
\\
& \multicolumn{4}{c}{LTE}   & & \multicolumn{5}{c}{LTE$_{\mathrm{corr}}$} \\
\cline{2-5} \cline{7-10}
\colhead{Star ID} & \colhead{\teff} & \colhead{\logg} & \colhead{\vt} & \colhead{\feh} & & \colhead{\teff} & \colhead{\logg} & \colhead{\vt}  & \colhead{\feh} & \colhead{$\sigma_{\mathrm{[Fe I /H]}}$} & \colhead{N$_\mathrm{\fei}$} & \colhead{N$_\mathrm{\feii}$} \\
 & \colhead{(K)} & & \colhead{(\kms)} & & & \colhead{(K)} & & \colhead{(\kms)}  &  & \colhead{(dex)} &}

\startdata                                                         
2MASS J00125284+4726278 &4400 &0.03 &1.91 &$-$2.68 & &4630 &0.82 &2.01 &$-$2.50 &0.23 &77 &11\\  
2MASS J01171437+2911580 &4414 &0.12 &1.70 &$-$2.72 & &4643 &0.88 &1.58 &$-$2.60 &0.21 &92 &14\\  
2MASS J01261714+2620558 &4875 &2.74 &1.21 &$-$0.93 & &5057 &3.28 &1.33 &$-$0.76 &0.17 &53 &8\\  
2MASS J02462013$-$1518418 &4700 &0.99 &1.83 &$-$3.10 & &4900 &1.45 &1.95 &$-$2.90 &0.13 &83 &10\\  
2MASS J04051243+2141326 &5396 &1.60 &2.03 &$-$2.80 & &5526 &1.91 &1.88 &$-$2.67 &0.18 &45 &10\\  
2MASS J04464970+2124561 &5889 &3.37 &2.12 &$-$1.96 & &5970 &3.49 &2.03 &$-$1.88 &0.24 &57 &10\\  
2MASS J05455436+4420133 &4358 &0.41 &1.54 &$-$2.91 & &4592 &1.24 &1.66 &$-$2.68 &0.23 &77 &12\\  
2MASS J06114434+1151292 &4270 &0.02 &1.58 &$-$2.95 & &4513 &0.78 &1.69 &$-$2.72 &0.18 &71 &9\\  
2MASS J06321853+3547202 &4877 &1.45 &0.90 &$-$2.95 & &5059 &1.80 &1.06 &$-$2.80 &0.18 &79 &7\\  
2MASS J07424682+3533180 &4705 &1.36 &0.95 &$-$2.90 & &4904 &1.70 &1.09 &$-$2.79 &0.15 &60 &7\\ 
2MASS J07532819+2350207 &5276 &2.10 &0.71 &$-$2.85 & &5418 &2.30 &0.92 &$-$2.88 &0.20 &36 &3\\  
2MASS J08011752+4530033 &4780 &1.59 &1.19 &$-$3.02 & &4972 &1.95 &1.32 &$-$2.90 &0.19 &89 &10\\  
2MASS J08203890+3619470 &4327 &0.66 &1.48 &$-$2.69 & &4564 &1.46 &1.54 &$-$2.51 &0.26 &50 &7\\ 
2MASS J08471988+3209297 &4403 &1.06 &2.23 &$-$2.47 & &4633 &1.59 &2.09 &$-$2.30 &0.21 &54 &7\\  
2MASS J09092839+1704521 &4709 &1.63 &1.10 &$-$2.41 & &4914 &2.15 &1.24 &$-$2.30 &0.11 &46 &4\\  
2MASS J09143307+2351544 &4311 &0.55 &1.40 &$-$3.45 & &4550 &1.15 &1.51 &$-$3.25 &0.21 &71 &12\\  
2MASS J09185208+5107215 &4890 &1.00 &1.44 &$-$3.18 & &5071 &1.39 &1.57 &$-$3.11 &0.20 &67 &7\\ 
2MASS J09261148+1802142 &4324 &0.67 &1.84 &$-$2.97 & &4562 &1.20 &1.71 &$-$2.70 &0.25 &73 &7\\  
2MASS J09563630+5953170 &4205 &0.65 &2.85 &$-$2.53 & &4454 &1.15 &2.77 &$-$2.22 &0.21 &54 &3\\  
2MASS J10122279+2716094 &4484 &0.75 &1.31 &$-$2.52 & &4706 &1.32 &1.43 &$-$2.46 &0.24 &70 &11\\
2MASS J10542923+2056561 &5029 &3.22 &1.08 &$-$0.71 & &5196 &3.58 &3.12 &$-$0.55 &0.19 &45 &6\\  
2MASS J11052721+3305150 &5121 &2.24 &1.17 &$-$3.12 & &5279 &2.48 &1.03 &$-$3.00 &0.24 &54 &4\\  
2MASS J12131230+2506598 &4605 &0.70 &1.40 &$-$2.98 & &4814 &0.95 &1.57 &$-$2.90 &0.26 &47 &8\\  
2MASS J12334194+1952177 &4140 &0.08 &1.38 &$-$3.10 & &4396 &0.58 &1.44 &$-$3.00 &0.26 &75 &8\\  
2MASS J12445815+5820391 &4173 &0.10 &1.57 &$-$2.93 & &4425 &0.90 &1.63 &$-$2.85 &0.18 &63 &7\\  
2MASS J13281307+5503080 &5700 &3.92 &0.49 &$-$0.19 & &5800 &4.13 &3.81 &$-$0.14 &0.18 &41 &5\\
2MASS J13525684+2243314 &4503 &1.37 &1.91 &$-$2.70 & &4723 &1.95 &1.52 &$-$2.55 &0.19 &50 &4\\  
2MASS J13545109+3820077 &4410 &0.25 &2.24 &$-$2.83 & &4639 &0.85 &2.11 &$-$2.70 &0.14 &61 &7\\  
2MASS J14245543+2707241 &5900 &3.72 &3.02 &$-$1.74 & &5980 &3.95 &2.87 &$-$1.70 &0.22 &32 &4\\  
2MASS J14445238+4038527 &4729 &1.79 &3.39 &$-$2.56 & &4926 &2.20 &1.81 &$-$2.45 &0.23 &34 &3\\  
2MASS J15442141+5735135 &4404 &0.26 &1.80 &$-$2.88 & &4550 &1.15 &1.71 &$-$2.75 &0.21 &78 &9\\  
2MASS J16374570+3230413 &4995 &1.48 &1.57 &$-$2.57 & &5165 &1.76 &1.61 &$-$2.45 &0.22 &82 &11\\  
2MASS J16380702+4059136 &5138 &2.90 &1.02 &$-$2.63 & &5294 &3.25 &2.74 &$-$2.50 &0.27 &51 &7\\  
2MASS J16393877+3616077 &5751 &2.45 &0.66 &$-$1.93 & &5846 &2.61 &2.33 &$-$1.95 &0.26 &47 &6\\  
2MASS J16451495+4357120 &4598 &0.98 &2.16 &$-$2.88 & &4808 &1.40 &1.04 &$-$2.71 &0.18 &61 &8\\  
2MASS J17041197+1626552 &4728 &0.97 &3.46 &$-$2.79 & &4925 &1.50 &1.13 &$-$2.66 &0.21 &46 &8\\  
2MASS J17045729+3720576 &4628 &2.10 &2.12 &$-$2.54 & &4835 &2.55 &2.01 &$-$2.45 &0.29 &38 &8\\  
2MASS J17125701+4432051 &4320 &0.17 &1.94 &$-$2.87 & &4558 &0.78 &2.11 &$-$2.70 &0.17 &77 &8\\  
2MASS J21463220+2456393 &5786 &2.80 &2.00 &$-$1.05 & &5685 &2.98 &1.85 &$-$1.10 &0.18 &44 &5\\  
2MASS J22175058+2104371 &5020 &1.47 &0.89 &$-$3.16 & &5188 &1.66 &0.95 &$-$3.07 &0.28 &43 &7\\  
2MASS J22424551+2720245 &4798 &1.56 &1.12 &$-$3.45 & &4988 &1.98 &1.12 &$-$3.30 &0.18 &68 &7\\ 
\enddata
\end{deluxetable*}

\section{Elemental Abundances}\label{sec:abund}

We could derive abundances, or at least meaningful upper limits, for the light, $\alpha$-, Fe-peak, and neutron-capture elements for all the target stars, including C, O, Na, Mg, Al, Si, Ca, Sc, Ti I, Ti II, V I, V II, Cr I, Cr II, Mn, Co, Ni, Cu, Zn, Sr, Ba, and Eu, using \texttt{MOOG} in \texttt{SMHr}. We measured the EWs of the absorption lines present in the spectra, and considered lines having EW $\leq 150$ m{\AA} and reduced equivalent widths (REWs)$\leq -4.5$ whenever possible, since they are on the linear part of the curve of growth, and are relatively insensitive to the choice of microturbulence.

Linelists, along with the isotopic and hyperfine structure for relevant elements including the neutron-capture elements, are obtained from the RPA standard linelists with updated $\log gf$ values \citep{roederer2018}, generated with \texttt{linemake}\footnote{https://github.com/vmplacco/linemake} \citep{linemake_vini}.Solar photospheric abundances have been used for the elements discussed in this study and are taken from \citet{asplund2009}.

\begin{figure*}
\centering
\includegraphics[width=2.0\columnwidth]{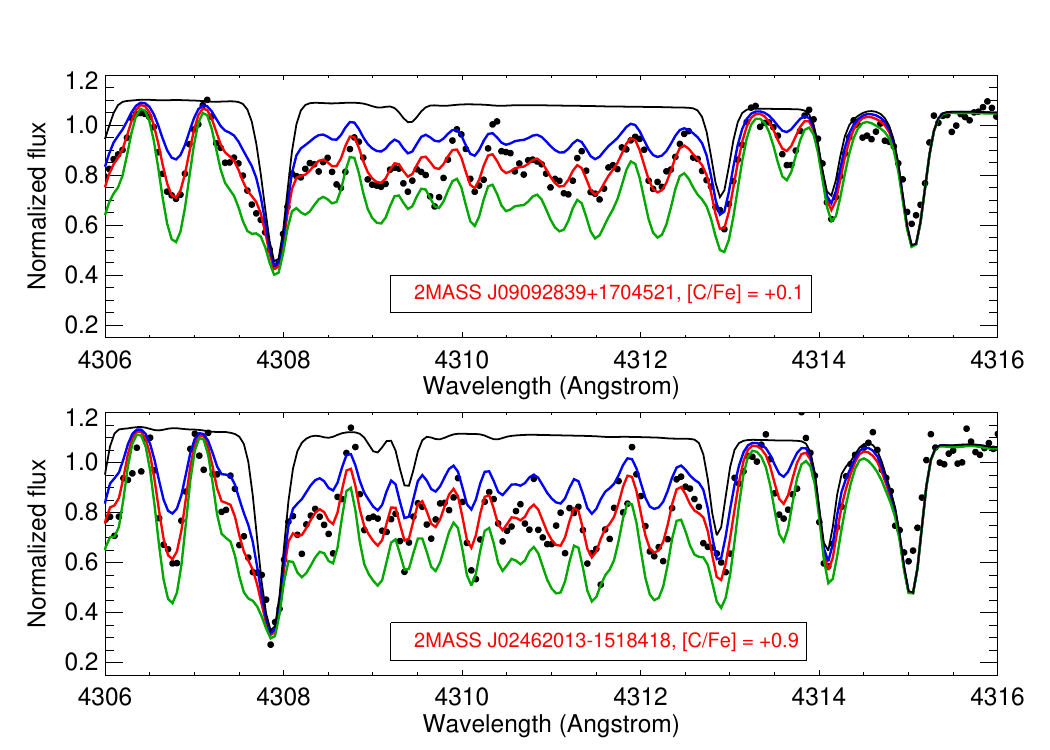}
\caption{Example spectral synthesis for the region of the molecular CH $G$-band. The red line shows the best-fit synthetic spectrum to the data (black dots). The blue and green lines mark deviations by $\pm 0.25$\,dex. The black line corresponds to the absence of carbon.}
\label{chsyn}
\end{figure*}

The trends for all the elements from C to Zn, and example spectral syntheses of the key lines, are shown in Figures \ref{chsyn}-\ref{ncap}. We also compare our results with those of metal-poor stars from \citet{jinabase2018}, including 
\citet{roederer2014b} and \citet{yong2013}, as well as previous RPA data releases from \cite{rpa1}, \citet{rpa2}, \citet{rpa3}, and \citet{rpa4}, as shown with gray-filled circles. 

\subsection{Carbon}

Carbon is an important element in studies of metal-poor stars, as it can be synthesized via multiple pathways in massive stars \citep{Liang2001,Farmer2021} and early supernovae \citep{bonifacio2015,Chan2020}, with implications for our understanding of the early Galaxy. Carbon is also produced by low- and intermediate mass AGB stars \citep{lugaro2003,Karlug2016}. Overall, it plays a key role in classifying the various stellar populations. Abundances for carbon in our sample stars were estimated by fitting the molecular CH $G$-band at 4315\,{\AA} via spectrum synthesis, as shown in Figure \ref{chsyn} for one C-normal star and one carbon-enhanced metal-poor (CEMP; [C/Fe] $\geq$ +0.7; \citealt{Beers2005,aoki2007apj}) star. The intensity of the molecular band is impacted by the assumed oxygen abundance, which in turn affects the amount of carbon that is locked into CO. Due to limitations in the available spectra, accurately determining the oxygen abundance is not feasible. We adopted [O/Fe] = +0.60 for the stars, consistent with observations of Milky Way halo stars with similar metallicities as discussed in \citep{amarsi19, asa24}. This assumption is supported by empirical observations of metal-poor stars in the Galactic halo and globular clusters, where [O/Fe] ratios range from +0.40 to +0.80, with +0.60 as a representative average \citep{Fulbright2003, Ramirez2012}. Consistent [O/Fe] ratios across various stellar populations reinforce our use of a uniform [O/Fe] ratio. Sensitivity tests show minimal impact from minor deviations around +0.60 on derived abundance patterns.

\begin{figure}
\centering
\hspace*{-0.8cm}
\includegraphics[scale=0.5]{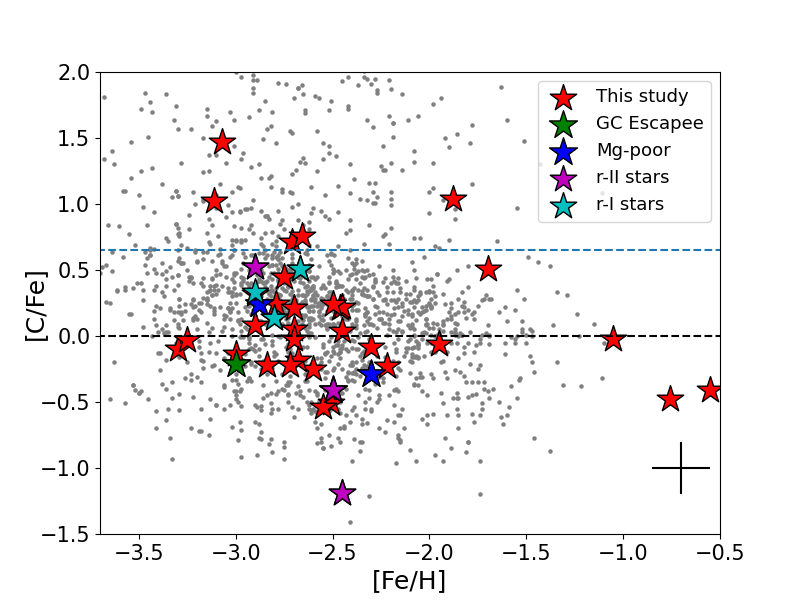}
\caption{Distribution of carbon, as a function of metallicity, [Fe/H]. The red stars denote the abundances of the stars in this study. The sample of stars from JINABASE \citep{jinabase2018}, including \citet{roederer2014b}, \citet{yong2013}, \citet{rpa1}, \citet{rpa2}, \citet{rpa3}, and \citet{rpa4}, are shown with gray dots. The blue-dashed line indicates the level above which stars are considered to be CEMP stars. The  Mg-poor VMP stars, and the globular cluster escapee are marked in blue and green, respectively. Typical error bars are indicated at the bottom right in the panel.}
\label{light}
\end{figure}

The range of C abundance ratios for the sample stars varies from [C/Fe] = $-$0.60 to [C/Fe] = $+$1.50, as seen in Figure \ref{light}. The C abundances are listed in Table \ref{tab:r_process_classification}. 
Since the majority of the stars are red giants,  corrections to the measured carbon abundances due to evolutionary effects have been computed following \citet{placco2014}, and incorporated in the reported abundances. Six stars are CEMP stars and they are among the most metal-poor stars in the sample. We also note the extremely low C abundance of the $r$-II star 2MASS J17045729+3720576, with [C/Fe] $\leq -1.2$ and [Fe/H] = $-$2.45, making it an interesting target for follow-up studies.

\subsection{Light Elements}

\begin{figure*}
\centering
\includegraphics[scale=0.5]{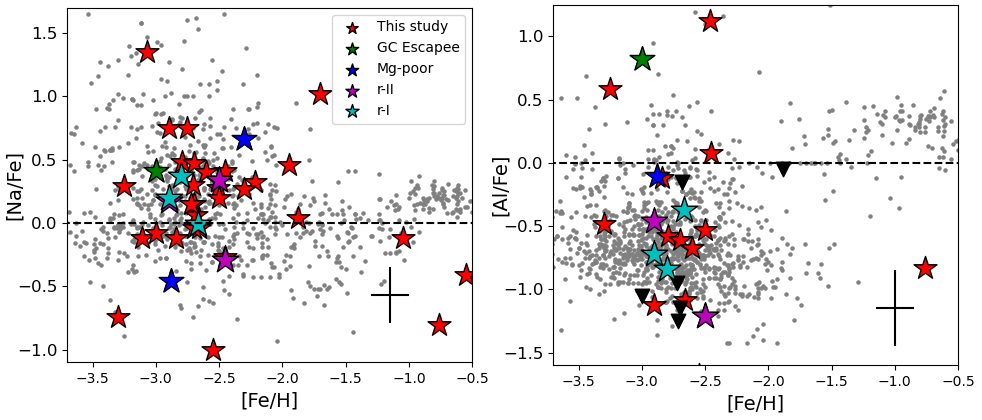}
\caption{Distribution of the odd-Z light elements Na and Al, as a function of metallicity, [Fe/H]. The red stars denote the LTE abundances of the stars in this study; filled downward-black triangles represent the derived upper limits. The individual elements are marked on the panels. The sample of stars from JINABASE \citep{jinabase2018}, including \citet{roederer2014b}, \citet{yong2013}, \citet{rpa2}, and \citet{rpa3}, are shown with gray dots. The  Mg-poor VMP stars, and the globular cluster escapee are marked in blue and green, respectively. Typical error bars are indicated at the bottom right in each panel.}
\label{oddz}
\end{figure*}

The odd-Z elements sodium and aluminium are mostly synthesized during hydrogen burning in the Ne-Na cycle \citep{cristal2015}, and via hydrostatic carbon and neon burning in massive stars \citep{nomoto2013}. In this study, the Na abundances are derived from the \nai\ doublet D1 and D2 at 5895\,{\AA} and 5889\,{\AA}. Non-LTE (NLTE) corrections for Na have been computed by \cite{andrievskyna} and \cite{lind2011}, and are around $-$0.10\,dex, but they can increase to $-$0.20\,dex depending on the logg values for a given metallicity regime. The abundances for Na are corrected by $-$0.15\,dex to account for the well-known NLTE effects. The final Na distribution is shown in Figure \ref{oddz}. The LTE abundances are listed in Table \ref{tab:abund_light}. The NLTE corrections are based on the average values for the given metallicity of the stars and slight deviations do not affect the final results.

Al abundances are estimated from the \ali\ resonance lines at 3961\,{\AA}. Due to the poor SNR of the fainter stars, we could only measure Al for 16 out of the 41 stars. The 1-D NLTE corrections were calculated from the calculations provided by \citet{nordlander2017}. We applied the corrections based on each star's $T_{\text{eff}}$ and $\log g$, ranging from +0.5 dex for stars at the base of the RGB to +1.1 dex for the coolest giants. The LTE abundances are listed in Table \ref{tab:abund_light}. We employed spectral synthesis for CEMP stars to account for blending of the Al line with CH.

Among the light elements, we find a large scatter for Al, as shown in the bottom panel of Figure \ref{oddz}. The Na and Al abundances have been corrected for NLTE in this figure. The scatter can partly be attributed to the larger uncertainties due to the poor SNR in the blue region of the spectra. Both Na and Al appear to follow the general trend found for metal-poor stars, with Na showing a larger scatter while Al is mostly sub-Solar towards the metal-poor end.

\begin{figure*}
\centering
\includegraphics[scale=0.7]{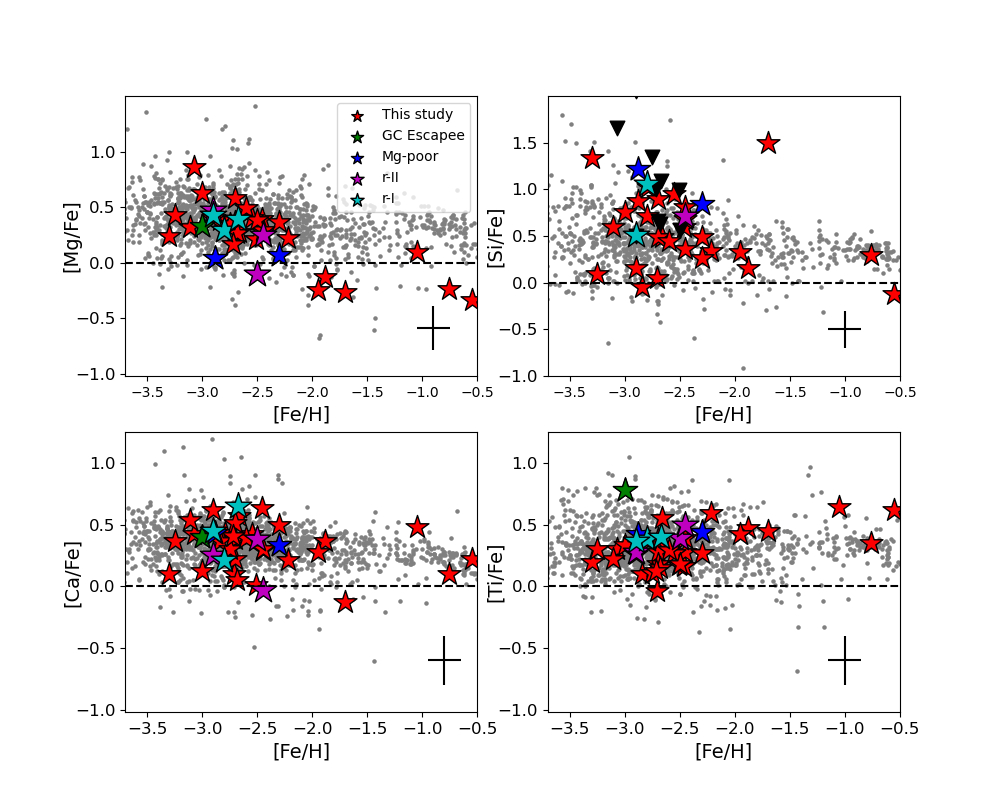}
\caption{Distribution of the $\alpha$-elements Mg, Si, Ca, and Ti, as a function of metallicity, [Fe/H]. The symbols and sources of literature data are the same as in Figure \ref{oddz}. Typical error bars are indicated at the bottom right in each panel. }
\label{alpha}
\end{figure*}

\begin{figure*}
\centering
\hspace*{-1.5cm}
\includegraphics[scale=0.7]{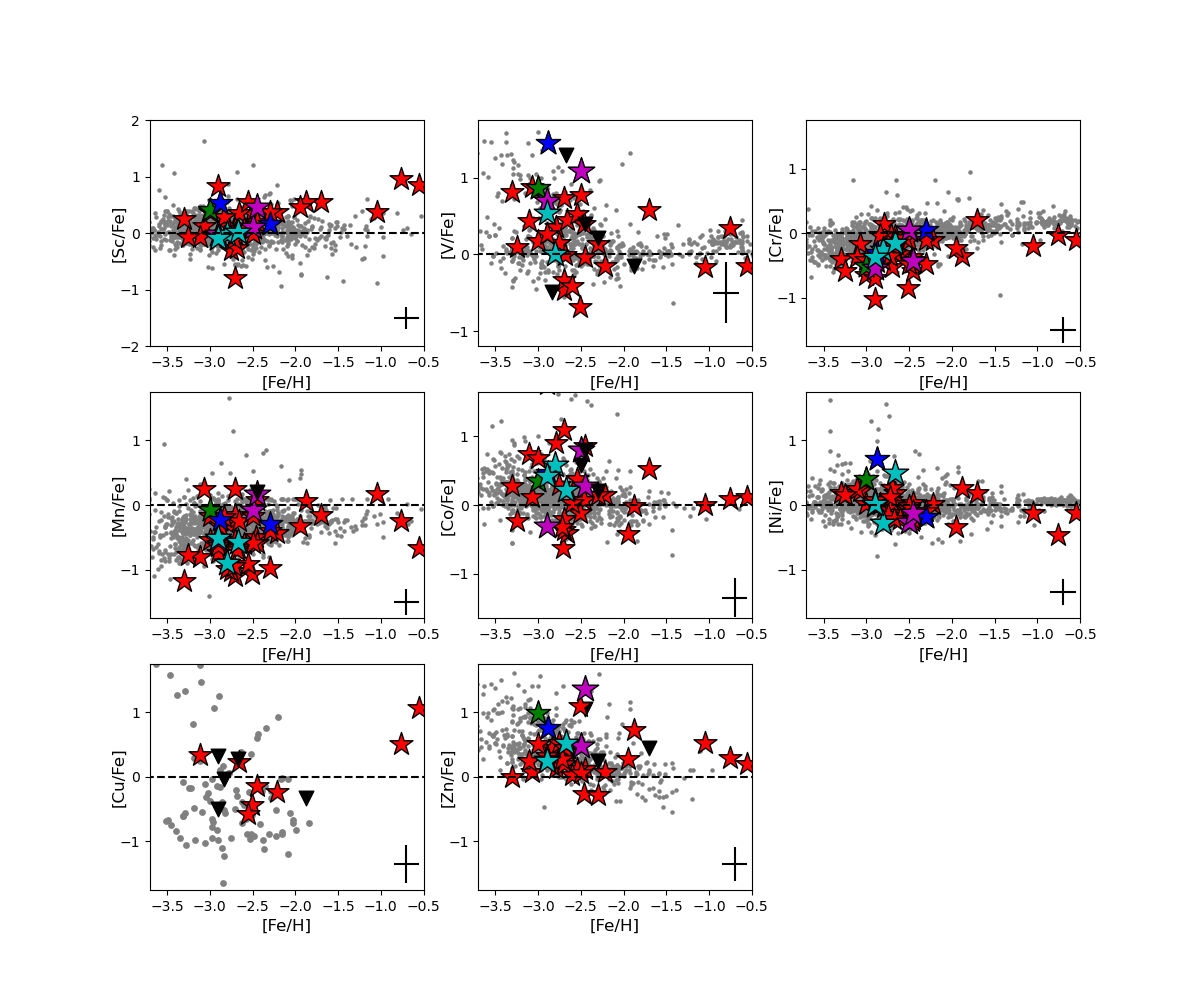}
\caption{Distribution of the Fe-peak elements formed by incomplete Si burning, as a function of metallicity, [Fe/H]. The symbols are the same as in Figure \ref{light}. Typical error bars are indicated at the bottom right in each panel.}
\label{fepeak1}
\end{figure*}

\begin{figure*}
\centering
\hspace*{-1cm}
\includegraphics[scale=1.1]{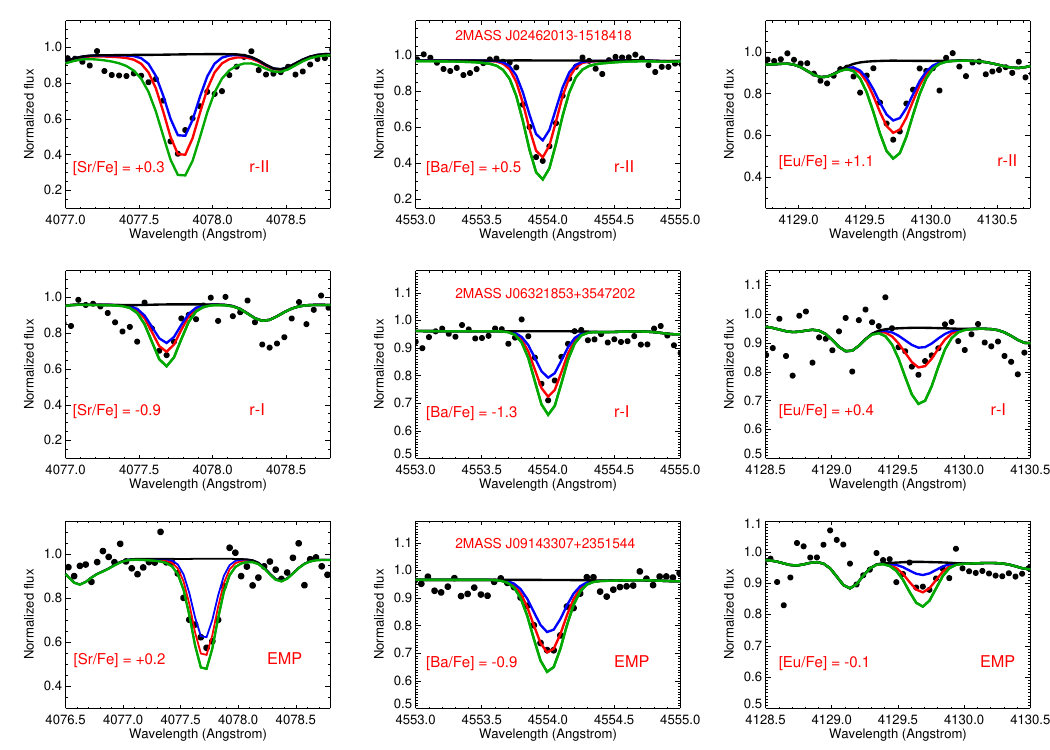}
\caption{Example spectral syntheses for lines of key $r$-process elements Sr, Ba, and Eu for 
2MASS J02462013$-$1518418, 2MASS J06321853+3547202,  and 2MASS J09143307+231544, which are $r$-II, $r$-I, and EMP stars, respectively. The red line indicates the best-fit synthetic spectrum; blue and green lines mark deviations by $\pm 0.20$\,dex. The black line corresponds to the absence of the given element.}
\label{rpsyn}
\end{figure*}

\begin{figure*}
\centering
\includegraphics[width=2.00\columnwidth]{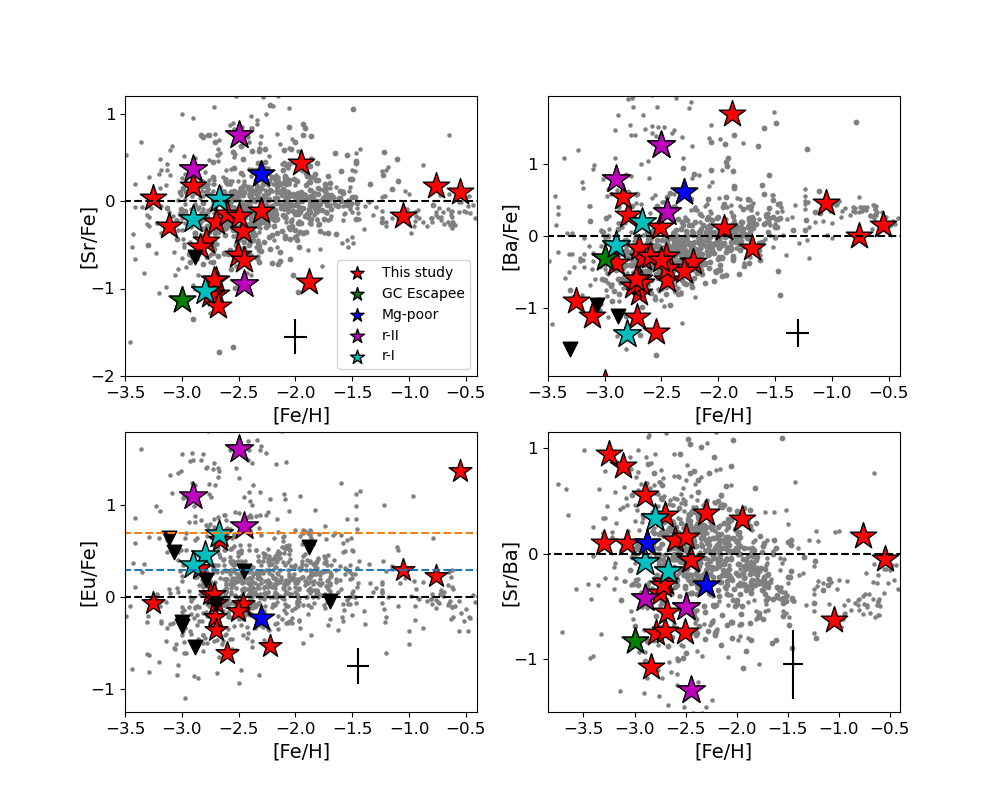}
\caption{Distribution of the neutron-capture element abundances for Sr, Ba, and Eu, and the ratio [Sr/Ba], as a function of metallicity,  [Fe/H]. The symbols are the same as in Figure \ref{light}. The error bars are indicated at the bottom-right in each panel. The red- and blue-dashed lines represent the limit for the $r$-II stars at [Eu/Fe] = +0.7, and $r$-I stars at [Eu/Fe] = +0.3, respectively.}
\label{ncap}
\end{figure*}

\begin{figure*}
\centering
\includegraphics[width=1.90\columnwidth]{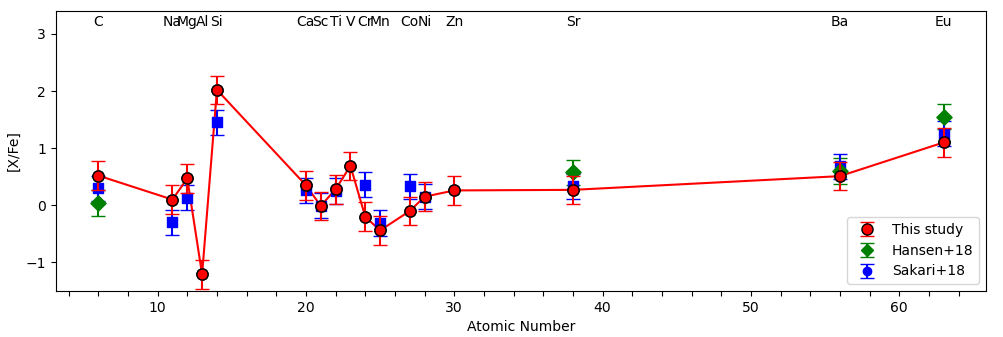}
\caption{Comparison of the derived abundances for 2MASS J02462013-1518418 from this study marked in red, and compared to the derived abundances from \citet{rpa1} and \citet{rpa2} in green and blue, respectively. }
\label{zigzag}
\end{figure*}

\subsection{The $\alpha$-elements}\label{sec:alpha-elem}

The $\alpha$-elements are produced in both the pre-explosive and explosive phases of CCSNe via several processes, such as carbon, oxygen, and neon burning \citep{heger2002, nomoto2013}. Transitions of oxygen are very limited in the optical domain. The forbidden [\oi]\ lines at 6300\,{\AA} and 6363\,{\AA} are largely dependent on gravity, and tend to be very weak, particularly in metal-poor stars. We attempted to measure the 6300\,{\AA} feature in our spectra, but because it is severely blended with telluric lines and suffers from considerable blends with Ni \citep{carlos2001}, no detections were achieved. 

Other $\alpha$-elements -- Mg, Si, Ca, and Ti -- could be measured for the program stars. For Mg, we refrained from using the transitions at 5172\,{\AA} and 5183\,{\AA}, as they were too strong for a reliable abundance estimation based on the EWs. We employed the reliable, clean Mg lines at 4167\,{\AA}, 4702\,{\AA}, and 5528\,{\AA}, which provide consistent estimates of Mg for the majority of the stars. For Si, we could not use the most prominent transition of Si at 3905\,{\AA} in many of the target stars due to the poor SNR, and thus employed the other weaker transition at 4102\,{\AA} wherever we could detect it in the spectra. Calcium is another very important indicator of $\alpha$-element abundances, and we could detect several transitions of Ca in the spectra for all the stars. We did not use the resonance line at 4216\,{\AA}, as it leads to systematically lower Ca abundances \citep{pinto2021}. We could detect several clean features of \tii\ and \tiii\ for all the stars in this study; the derived abundances are listed in Table \ref{tab:abund_light}.

Abundances for the $\alpha$-elements are shown in Figure \ref{alpha}. The metal-poor stars exhibit an elevated $\langle{\rm [Mg/Fe]}\rangle = +0.35$, as expected for halo stars \citep{Mashonkina2019}. However, the CEMP star 2MASS J22175058+2104371 has an over-abundance of [Mg/Fe] $= +0.86$; this has also been previously observed in other CEMP stars \citep{Aoki2002z}. For the metal-poor stars, their [Si/Fe] ratios show the usual enhancement of [$\alpha$/Fe] $\sim +0.4$, albeit with large scatter. There appears a gradual decrease with increasing [Fe/H] ([Fe/H] $\ge -1.0$), marking the beginning of contributions from Type Ia supernovae. The overall scatter tends to decrease with increasing  metallicity. The derived Ca abundance ratios varied between [Ca/Fe] =  0.0 and +0.70. The mean $\langle$[Ca/Fe]$\rangle$ = +0.36 for the sample stars is consistent with the typical halo $\alpha$-enhancement of [$\alpha$/Fe] = $+$0.40. The mean Ti abundances of $\langle$[Ti/Fe]$\rangle$ = +0.34 follows the $\alpha$-enhancement ratios of other halo stars. However, Ti abundances show comparably little scatter across the entire metallicity range of the program stars. Among the VMP/EMP stars, 2MASS J16380702+4059136 exhibits a slightly sub-Solar [Mg/Fe]$  = -0.03$, a difference of 0.38\,dex from the mean value. For several stars, the Mg abundances are at the solar levels which is also significantly lower than the usual $\alpha$-enhanced halo stars. We selected the VMP/EMP stars with lower Mg but normal Ca abundances and marked them in blue in Figure \ref{alpha}. These are interesting candidates for dedicated studies to understand the nature of the progenitor supernovae.

\subsection{Fe-peak Elements}

The vast majority of the Fe-peak elements found in metal-poor stars are synthesized by the incomplete (e.g., Cr and Mn) and complete (e.g., Co and Ni) combustion of silicon in Type II supernovae \citep{nakamura1999,nissen2023abundances}.  However, Type Ia supernovae can also contribute the Fe-peak elements, particularly in the case of stars at the metal-rich end of the present sample. The derived LTE abundances for all the detected Fe-peak elements are listed in Table \ref{tab:abund_fe_peak}, and the distribution is shown in Figure \ref{fepeak1}; the derived abundances have not been corrected for NLTE effects. The expected range of corrections are provided as a reference for the reader. Hyperfine structure (HFS) was taken into account for the Fe-peak elements Sc, V, Mn, and Co as necessary, and spectral synthesis was used to derive the abundances for those lines. 

Among the Fe-peak elements, Sc is produced by supernovae of varying mass ranges. Sc production in CCSNe peaks for progenitors around 20 M$_\odot$ (2.0 $\times$ 10$^{-5}$ M$_\odot$), varying from 1.0 $\times$ 10$^{-5}$ M$_\odot$ at 15 M$_\odot$ to 1.2 $\times$ 10$^{-5}$ M$_\odot$ at 30 M$_\odot$, influencing the chemical evolution of its natal sub-halo, which also depends on the IMF and star formation history \citep{Woosley1995, nomoto2013}. Scandium abundances were derived from multiple lines, with the transition at 4254\,{\AA} being the most prominent. The distributions of Fe-peak element abundances is shown in Figure \ref{fepeak1}. The derived Sc abundances of the sample stars exhibit a large scatter, indicating that the parent gas cloud had contributions from a wide range of supernovae masses \citep{Chieffi2002}. However, the trend for the [Sc/Fe] ratio stays mostly constant, with a slight increase towards the metal-rich end of our sample.  It was difficult to obtain clean \vii\ lines in the spectra, although we could detect \vi\ lines for many of the stars. The \vi\ lines are known to be strongly affected by NLTE effects \citep{bergemann2010}. The [V/Fe] ratio exhibits a large scatter, which decreases as metallicity increases, until around [Fe/H] $= -2.0$, after which it flattens out. 

Multiple \cri\ lines could be detected in the spectra, including the stronger ones at 4646\,{\AA} and 5206\,{\AA}. Derived abundances are known to suffer from large NLTE effects \citep{bergemann2010}. We could also measure \crii\ lines in some of the evolved stars. A mean difference of 0.25\,dex was obtained between the Cr I and Cr II lines in the current sample, consistent with previous studies (e.g., 
\citealt{bonifacio2009,Cowan20,Sneden2023}). The [Cr/Fe] ratio displays a very tight correlation with [Fe/H]; it slightly increases with increasing [Fe/H] at the lowest metallicities, and then remains roughly constant above [Fe/H] $ = -2.0$. The Mn abundances for most stars were derived by employing the resonance Mn triplet at 4030\,{\AA} and an additional line at 4823\,{\AA}. Other weaker features are taken into account only when the SNR is too low in the 4030\,{\AA} region to measure meaningful abundances. However, these lines are prone to 3D and NLTE corrections ranging from 0.3 to 0.6\,dex, as reported by \citet{bergemann2019}. We could not detect any Mn II lines in the spectra.  The [Mn/Fe] ratio also exhibits a large dispersion, with a slight increase in [Mn/Fe] with increasing metallicity. These trends for Cr and Mn have been reported for other samples of metal-poor stars \citep{cayrel2004,Lai2008,Ban2018,Sneden2023}. They may indicate deeper mass cuts in the progenitor supernovae, and a dependence of a neutron excess on metallicity \citep{heger2010,nomoto2013}.

Abundances for Co were mostly derived from the features at 3995\,{\AA} and 4121\,{\AA}; we have been able to at least measure upper limits for the sample stars.  The [Co/Fe] ratios in our sample stars exhibit a large dispersion, accompanied by a slight decrease with increasing metallicity. Cobalt is particularly over-produced relative to Fe in short-lived massive stars during the explosion. 
Ni is expected to track the Fe content. The mean abundance ratio for Ni is $\langle {\rm [Ni/Fe]}\rangle = +0.10$ for the sample. The observed scatter for Ni is significantly less-pronounced than for the other Fe-peak elements. The [Ni/Fe] ratio for our sample stars maintains a tight correlation with [Fe/H], which hardly varies over the entire metallicity range.This might be expected, as Ni and Fe are synthesized in the same region, and hence, it is very difficult to change the ratio \citep{kobayashi2020}.  However, we find that the EMP star 2MASS J07532819+2350207 has a very high ratio of [Ni/Fe] $ = +0.86$, accompanied by elevated Co and Zn abundances. Copper abundances could be derived for very few stars in our sample using the 5105.5\,{\AA} line. While typically classified as a Fe-peak element, it is noteworthy that significant quantities of Cu can also be synthesized through the weak $s$-process \citep{Pignatari2010,Nishimura2017}. The sample size is insufficient to derive a significant trend for Cu with metallicity. The [Cu/Fe] ratio varies between $+0.5$ and $-0.5$ for the metal-poor stars in the sample, consistent with previous studies.  Zinc is produced in the deepest layers of CCSNe, and is enhanced for HNe with higher explosion energy \citep{kobayashi2020}. Zn is an important element to constrain the mass range of the progenitor supernovae, and could be detected in the majority of our sample stars. We have employed the only two useful lines of Zn at 4722\,{\AA} and 4810\,{\AA} for determining the abundances.  The [Zn/Fe] ratios tend to decrease with increasing metallicity. This is also expected, as the yields of Zn decrease for less-massive supernovae at higher metallicites. 

\subsection{Neutron-capture Elements}

Among the neutron-capture elements, we could derive the abundances and meaningful upper limits for Sr, Ba, and Eu, as described below. Example syntheses of Sr, Ba, and Eu for $r$-I, $r$-II, and other EMP stars are shown in Figure \ref{rpsyn}. Best fits to key lines are shown with red lines. Deviations of $\pm 0.20$\,dex are marked by the blue and green lines.  The black lines indicate the absence of a given element. Based on the resulting abundances, we classify the objects as  limited-$r$ $r$-I, $r$-II, or EMP/VMP stars.

\subsubsection{Strontium}

Strontium has a complex nucleosynthesis history, but is largely contributed by the $r$-process at low metallicites.  The contribution from the $s$-process \citep{Lugaro2012,Karlug2016} rises as the metallicity increases (e.g most of the solar abundance of Sr comes from the s-process \citep{Prantzos2020}). For the $r$-process origin of Sr, it is hypothesized to be produced during the explosion phase of magneto-rotationally driven supernovae as well as in neutron star mergers \citep{Reichert20,Perego2022}. Indeed, it was the first neutron-capture element to have been detected in the neutron star merger GW170817 \citep{gw,watson2019,domoto22}. Strontium can also be produced via the $s$-process in intermediate-mass AGB stars and weak $s$-process in massive stars \citep{Pignatari2010,Norfolk19}. The abundances of Sr for our sample stars are derived by fitting the 4077\,{\AA} resonance line and the strong 4215\,{\AA} line, wherever possible, with spectrum synthesis. For a few stars, we noticed saturation of the 4077\,{\AA} feature or blending of the 4215\,{\AA} line. In those cases, the lines were not employed to derive the final abundances for those stars. The Sr abundances for our sample stars are listed in Table \ref{tab:r_process_classification}. 

\subsubsection{Barium}

Barium is one of the most widely studied species among the neutron-capture elements. 
It is produced by both the $s$-process in AGB stars via thermal pulsations as in the case for Sr \citep{Lugaro2012,Cristallo2015,Karlug2016,Hartogh2023} and the $r$-process via explosive events \citep{Duggan2018,Cescutti2021,Cowan2021}. It can also be produced in the $i$-process \citep{Choplin2021} and possible weak $s$-process in rotating massive stars \citep{Frischknecht16,Karlug2016}. However, the $r$-process is expected to be the dominant contributor at the lowest metallicites, with increasing contributions from the $s$-process with increasing metallicity \citep{Simmerer2004,magrini2018} over cosmic times. Barium abundances have been derived for our sample stars by spectrum fitting of three prominent features at 4554\,{\AA}, 4934\.{\AA}, and 6141\,{\AA}. The 4934\,{\AA} feature is more difficult to analyze, as it yields larger uncertainties due to  significant Fe blends in the line wings \citep{Gallagher2010}. As a result, this line is discarded whenever the derived abundances deviate strongly from the other two features. The $r$-process isotope ratios from \cite{sneden2008} were adopted for the spectral analysis. NLTE effects have been studied by \cite{korotin}, and are usually less than 0.1\,dex for the line at 4554\,{\AA} at the given metallicity. However, for most of the stars, we note a good convergence of the derived Ba abundances from all three features. The final measured values are listed in Table \ref{tab:r_process_classification}. 

\subsubsection{Europium}

At low metallicites, Eu is exclusively produced by the $r$-process, and NSMs are expected to be one of the primary sites for production of Eu \citep{cain2018,cote2018,erikansm}. Even at Solar metallicities, the majority of Eu is produced by the $r$-process \citep{Bartos2019,Schatz2022}. We derived the abundances using spectral synthesis of the line at 4129\,{\AA}, which is the strongest Eu line in the observed wavelength range. We also detected a weaker line at 4205\,{\AA} in a few stars, but most of the final abundances listed in Table \ref{tab:r_process_classification} are based on the feature at 4129\,{\AA}. However, due to the low SNR of the spectra, this line could not be measured for all of the stars; meaningful upper limits could be obtained in these cases.

The trends for the neutron-capture elements, and the [Sr/Ba] ratio, as functions of metallicity, are shown in Figure \ref{ncap}. They exhibit similarities in the individual trends with metallicity. The majority of our sample stars exhibit sub-Solar Sr and Ba abundances at low metallicities, which slightly increases and stabilizes around [Sr/Fe, Ba/Fe] = 0.0 at metallicites higher than [Fe/H] $ = -2.0$.

The levels of Eu used to define the $r$-I ($+0.3 < {\rm [Eu/Fe]} \leq +0.7$)  and $r$-II  ([Eu/Fe] $> +0.7$) stars by \citet{rpa4} are shown with dashed lines in the figure. The $r$-I and $r$-II stars are also defined to have [Ba/Eu] $< 0$, which applies for most of the current sample of stars with enhanced Eu, as shown in Table \ref{tab:r_process_classification}. However, we also note the presence of several interlopers with $s$-process or $r/s$-process dominance, identified by [Ba/Eu] $> 0$ in the sample.

\subsection{Uncertainties}\label{sec:uncert}

Abundance uncertainties can be attributed to two primary sources: the SNRs of the observed spectra and associated quality of the line fit, and the uncertainties in the derived stellar parameters. To assess the impact of the SNR, we employ Equation 6 from \citet{Cayrel88} to estimate the associated uncertainties in our abundance determinations as outlined in \citet{banli}. The computed uncertainties based on the SNR are on the lower side, typically less than 0.1 dex. The stellar parameters are known to suffer from systematic uncertainties, which are computed as outlined in \citet{ji2016b} and \citet{rpa3}. These uncertainties are estimated for variations of 150\,K in \teff, 0.25\,dex in log $g$, and 0.2 km s$^{-1}$ in microturbulent velocity. Additionally, we calculate the uncertainties in metallicity ([Fe/H]) based on the standard deviations in the abundances of Fe I and Fe II lines. As an example, these systematic uncertainties for 2MASS J02462013$-$1518418 are provided in Table \ref{tbl:sysA}. The  uncertainties are combined in quadrature to determine the total systematic uncertainty in the abundance measurements.

\subsection{Comparison with Previous RPA Studies}

To further test the accuracy of our derived abundances, we compare the abundances of one star, 2MASS J02462013-1518418, with those derived previously by \citet{rpa1} and \citet{rpa2}.
Figure \ref{zigzag} shows the comparison. \citet{rpa1}  derived abundances for the elements C, Sr, Ba, and Eu, while for \citet{rpa2}, we were able to compare our abundances with the $\alpha$- and Fe-peak elements in common. Our  abundances of most elements agree within reported uncertainties (see Section\,\ref{sec:uncert}). However, we notice larger deviations for Si from the abundances derived by \citet{rpa2}, which is due to using different lines in the spectrum.

\begin{deluxetable}{lrrrr}
\tablecolumns{5}
\tablewidth{0pt}
\tablecaption{Example Systematic Uncertainties for 2MASS J02462013$-$1518418 \label{tbl:sysA}}
\tablehead{
\colhead{Element} & \colhead{$\Delta${T$_{\rm eff}$}} & \colhead{$\Delta${\logg}} & \colhead{$\Delta${\micro}} & \colhead{Total}\\ 
\colhead{} & \colhead{(dex)} & \colhead{(dex)} & \colhead {(dex)} & \colhead{(dex)}}
\startdata
CH (syn)   & $+$0.19 & $-$0.07 & $-$0.02 &  0.20\\
Na I       & $+$0.14 & $-$0.05 & $-$0.07 &  0.16\\
Mg I       & $+$0.11 & $-$0.07 & $-$0.02 &  0.13\\
Al I       & $+$0.15 & $-$0.09 & $-$0.09 &  0.20\\
Si I       & $+$0.11 & $-$0.03 & $-$0.04 &  0.12\\
Ca I       & $+$0.13 & $-$0.04 & $-$0.07 &  0.15\\
Sc II (syn)& $+$0.10 & $+$0.05 & $-$0.06 &  0.13\\
Ti I       & $+$0.19 & $-$0.02 & $-$0.03 &  0.19\\
Ti II      & $+$0.08 & $+$0.06 & $-$0.05 &  0.11\\
V II       & $+$0.09 & $+$0.04 & $-$0.06 &  0.12\\
Cr I       & $+$0.19 & $-$0.03 & $-$0.08 &  0.21\\
Mn I       & $+$0.28 & $-$0.04 & $-$0.04 &  0.29\\
Co I (syn) & $+$0.20 & $-$0.07 & $-$0.03 &  0.21\\
Ni I       & $+$0.16 & $-$0.05 & $-$0.08 &  0.19\\
Sr II (syn)& $+$0.18 & $+$0.08 & $-$0.09 &  0.22\\
Ba II (syn)& $+$0.17 & $+$0.09 & $-$0.10 &  0.22\\
Eu II (syn)& $+$0.13 & $+$0.11 & $+$0.06 &  0.18\\
\enddata
\end{deluxetable}

\section{Results and Discussion}\label{sec:disc}

\subsection{Classification of the Observed Stars}

Following the abundance analysis, our sample of stars have been classified following the standard schemes, as described in \citet{Beers2005} and \citet{rpa4}. The classifications are based on the derived abundances for the key elements, and are shown in Table \ref{tab:r_process_classification}. For classification, Mg abundances are adopted as an indicator of the $\alpha$-element abundances, while Sr, Ba, and Eu are important neutron-capture elements that are widely used to study the RPE stars. Accordingly, the current sample comprises 1 limited-$r$ star, 3 $r$-I stars, 4 $r$-II stars, 5 CEMP stars, 6 Mg-poor stars, and 23 VMP/EMP stars. We note that 2MASS J16380702+4059136 is an $r$-II star with a low Mg abundance; hence it falls under both classes.

\begin{table*}
\scriptsize
    \centering
    \caption{Select Abundances and Classification of Program Stars}
    \label{tab:r_process_classification}
    \begin{tabular}{lccccccccccccl} 
        \hline
       Star name & [Fe/H] & [C/Fe]$^{o}$ & $\Delta$[C/Fe]$^{a}_{corr}$ &[Mg/Fe]& [Sr/Fe] & [Ba/Fe] & [Eu/Fe] & [Ba/Eu] &Classification\\
		\hline
2MASS J00125284+4726278  &$-$2.50 &$-$0.52 &0.76 &+0.38   &$-$0.07$\pm$0.17 &$-$0.33$\pm$0.16         &\dots &\dots &VMP \\
2MASS J01171437+2911580  &$-$2.60 &$-$1.02 &0.77 &+0.49   &$-$0.05$\pm$0.19 &$-$0.28$\pm$0.21         &$-$0.61$\pm$0.18 &+0.33 &VMP\\
2MASS J01261714+2620558  &$-$0.76 &$-$0.50 &0.02 &$-$0.24 &+0.27$\pm$0.16   &\phantom{+}0.00$\pm$0.16 &+0.23$\pm$0.18 &$-$0.23 &MP\\
2MASS J02462013$-$1518418&$-$2.90 &+0.10   &0.42 &+0.47   &+0.30$\pm$0.18   &+0.79$\pm$0.20           &+1.10$\pm$0.24 &$-$0.31 &$r$-II\\
2MASS J04051243+2141326  &$-$2.67 &+0.16   &0.35 &+0.39   &+0.18$\pm$0.22   &+0.19$\pm$0.17           &+0.70$\pm$0.24 &$-$0.51 &$r$-I\\ 
2MASS J04464970+2124561  &$-$1.88 &+1.04   &0.00 &$-$0.13 &$-$0.83$\pm$0.16 &+1.70$\pm$0.22           &$<$+0.55 &\dots &Mg-poor, CEMP-s\\
2MASS J05455436+4420133  &$-$2.68 &$-$0.94 &0.76 &+0.27   &$-$1.10$\pm$0.25 &$-$0.65$\pm$0.26        &\dots &\dots &VMP\\ 
2MASS J06114434+1151292  &$-$2.72 &$-$0.98 &0.76 &+0.17   &$-$0.80$\pm$0.21&$-$0.60 $\pm$0.22       &+0.03$\pm$0.18 &$-$0.63 &Mg-poor, VMP\\ 
2MASS J06321853+3547202  &$-$2.80 &$-$0.39 &0.53 &+0.29   &$-$0.93$\pm$0.25&$-$1.37$\pm$0.22        &+0.46$\pm$0.19 &$-$1.83 &$r$-I, VMP\\
2MASS J07424682+3533180  &$-$2.79 &$-$0.30 &0.54 &+0.33   &$-$0.36$\pm$0.16 &+0.29$\pm$0.16          &$<$+0.19 &\dots &VMP\\
2MASS J07532819+2350207  &$-$2.88 &+0.23   &0.01 &+0.04   &$<-$0.64   &$<-$1.11          &$<-$0.54 &\dots &Mg-poor, VMP\\
2MASS J08011752+4530033  &$-$2.98 &$-$0.01 &0.33 &+0.43   &$-$0.11$\pm$0.16 &$-$0.13$\pm$0.20        &+0.35$\pm$0.16 &$-$0.48 &$r$-I, EMP\\
2MASS J08203890+3619470  &$-$2.51 &$-$1.28 &0.77 &+0.21   &$-$0.52$\pm$0.22 &+0.12$\pm$0.23          &$-$0.15$\pm$0.127 &+0.27 &VMP\\
2MASS J08471988+3209297  &$-$2.30 &$-$1.04 &0.75 &+0.07   &+0.41$\pm$0.16   &+0.61$\pm$0.16          &$-$0.23$\pm$0.17 &+0.84 &Mg-poor, VMP\\
2MASS J09092839+1704521  &$-$2.30 &$-$0.47 &0.39 &+0.37   &$-$0.01$\pm$0.17 &$-$0.49$\pm$0.19        &\dots &\dots &VMP\\
2MASS J09143307+2351544  &$-$3.25 &$-$0.75 &0.72 &+0.43   &+0.17$\pm$0.20   &$-$0.91$\pm$0.16        &$-$0.06$\pm$0.22 &$-$0.85 &Limited-r, EMP\\
2MASS J09185208+5107215  &$-$3.11 &+0.31   &0.71 &+0.32   &$-$0.19$\pm$0.21 &$-$1.12$\pm$0.24        &$<$+0.65 &\dots &CEMP-no\\
2MASS J09261148+1802142  &$-$2.70 &$-$0.71 &0.76 &+0.37   &$-$0.96$\pm$0.15 &$-$0.79$\pm$0.16       &$<-$0.04 &\dots &VMP\\
2MASS J09563630+5953170  &$-$2.22 &$-$1.01 &0.78 &+0.22   &\dots   &$-$0.36$\pm$0.16        &$-$0.53$\pm$0.16 &+0.17 &VMP\\
2MASS J10122279+2716094  &$-$2.46 &$-$0.56 &0.77 &+0.39   &$-$0.24$\pm$0.19 &$-$0.28$\pm$0.18        &$-$0.07$\pm$0.16 &$-$0.21 &VMP\\
2MASS J10542923+2056561  &$-$0.55 &$-$0.43 &0.02 &$-$0.34 &+0.20$\pm$0.21    &+0.15$\pm$0.22          &+1.38$\pm$0.19 &$-$1.23 &$r$-II, MP\\
2MASS J11052721+3305150  &$-$3.00 &$-$0.21 &0.00 &+0.33  &$-$1.03$\pm$0.18  &$-$0.30$\pm$0.16       &$<-$0.27 &\dots &EMP, GCE\\
2MASS J12131230+2506598  &$-$2.90 &$-$0.67 &0.75 &+0.40  &+0.27$\pm$0.19    &$-$0.38$\pm$0.26        &\dots &\dots &EMP\\
2MASS J12334194+1952177  &$-$3.00 &$-$0.89 &0.75 &+0.63  &\dots    &$-$2.05$\pm$0.22        &$<-$0.31 &\dots &EMP\\
2MASS J12445815+5820391  &$-$2.84 &$-$0.98 &0.76 &+0.38  &$-$0.42$\pm$0.23  &+0.55$\pm$0.21          &+0.32$\pm$0.17 &+0.23 &VMP\\
2MASS J13281307+5503080  &$-$0.14 &$-$0.22 &0.00 &$-$0.04 &$<-$0.81   &+0.46$\pm$0.16          &$<$+0.39 &\dots &MP\\
2MASS J13525684+2243314  &$-$2.55 &$-$1.09 &0.55 &+0.40  &\dots   &$-$1.34$\pm$0.20         &\dots &\dots &VMP\\
2MASS J13545109+3820077  &$-$2.70 &$-$0.55 &0.76 &+0.58  &$-$0.80$\pm$0.16 &$-$0.17$\pm$0.22         &$-$0.36$\pm$0.26 &+0.19 &VMP\\
2MASS J14245543+2707241  &$-$1.70 &+0.51   &0.00 &$-$0.26&\dots   &$-$0.17$\pm$0.19         &$<-$0.04 &\dots &Mg-poor, VMP\\
2MASS J14445238+4038527  &$-$2.45 &$-$0.16 &0.20 &+0.29  &\dots   &$-$0.44$\pm$0.18         &$<$+0.29 &\dots &VMP \\
2MASS J15442141+5735135  &$-$2.75 &$-$0.30 &0.75 &+0.35  &$-$0.97$\pm$0.22 &$-$0.70$\pm$0.16         &+0.04$\pm$0.26 &$-$0.74 &VMP\\
2MASS J16374570+3230413  &$-$2.45 &$-$0.26 &0.48 &+0.28  &$-$0.57$\pm$0.16 &$-$0.61$\pm$0.17         &\dots &\dots &VMP\\
2MASS J16380702+4059136  &$-$2.50 &$-$0.42 &0.01 &$-$0.10&+0.86$\pm$0.16   &+1.27$\pm$0.19           &+1.62$\pm$0.16 &$-$0.35 &$r$-II, Mg-poor\\
2MASS J16393877+3616077  &$-$1.95 &$-$0.07 &0.01 &$-$0.25&+0.54$\pm$0.17   &+0.11$\pm$0.15           &\dots &\dots &MP\\
2MASS J16451495+4357120  &$-$2.71 &$-$0.03 &0.74 &+0.28  &\dots   &$-$1.13$\pm$0.16         &$-$0.21$\pm$0.16 &$-$0.92 &CEMP-no\\
2MASS J17041197+1626552  &$-$2.66 &+0.03   &0.73 &+0.28  &\dots   &$-$0.31$\pm$0.22         &+0.62$\pm$0.16 &$-$0.93 &CEMP-no\\
2MASS J17045729+3720576  &$-$2.45 &$-$1.20 &0.01 &+0.25  &$-$0.85$\pm$0.18 &+0.34$\pm$0.16           &+0.80$\pm$0.19 &$-$0.46 &$r$-II, VMP\\
2MASS J17125701+4432051  &$-$2.70 &$-$0.78 &0.76 &+0.29  &$-$0.14$\pm$0.16 &$-$0.60$\pm$0.22         &$-$0.06$\pm$0.21 &$-$0.54 &VMP\\
2MASS J21463220+2456393  &$-$1.05 &$-$0.04 &0.02 &+0.10  &$-$0.07$\pm$0.20 &+0.46$\pm$0.23           &+0.30$\pm$0.16 &+0.16 &MP\\ 
2MASS J22175058+2104371  &$-$3.07 &+1.09   &0.38 &+0.86  &\dots   &$<-$0.96           &$<$+0.55 &\dots &Mg-rich, CEMP-no\\ 
2MASS J22424551+2720245  &$-$3.30 &$-$0.37 &0.27 &+0.24  &\dots   &$<-$1.58           &\dots &\dots &EMP\\ 
	\hline
	\end{tabular}
\tablenotetext{a}{Indicates correction for evolutionary effects from \citet{placco2014}. This value should be added to the ``as observed" [C/Fe] value in order to obtain the corrected [C/Fe] abundance.}
\tablenotetext{o}{Indicates derived abundance before correction.}
\end{table*}

\subsection{The $\alpha$- and Fe-peak Elements: Tracing the Sample Stars' Supernovae 
Progenitors}\label{alphaa}

\begin{figure*}
\centering
\includegraphics[width=2.10\columnwidth]{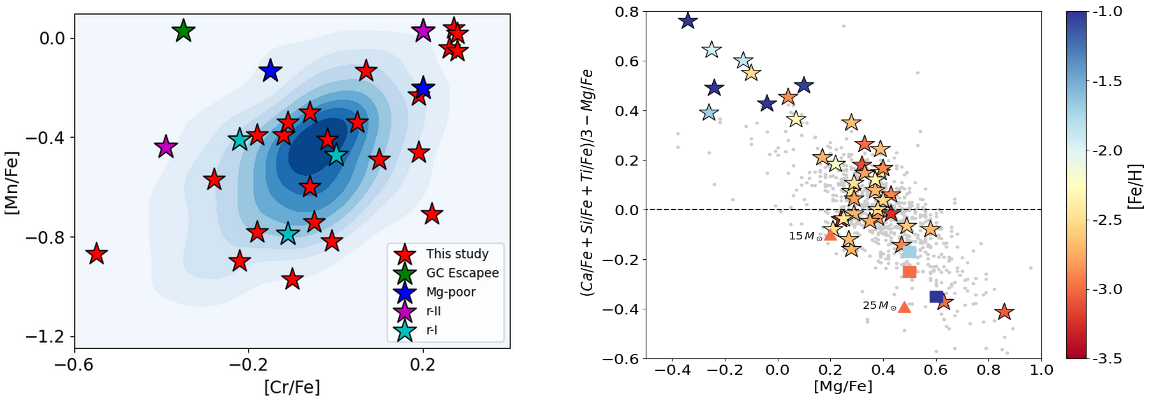}
\caption{Left panel: Distribution of [Mn/Fe] vs. [Cr/Fe] for our sample stars and stars from the literature, including \citet{roederer2014b}, \citet{rpa3}, and \citet{yong2013}, shown as a density plot. The darker colors indicate a higher density of points in the parameter space; lighter colors indicate a lower density of points. The symbols are the same as in previous figures. Right panel: Trend for $\rm \Delta\alpha = (([Ca/Fe]+[Si/Fe]+[Ti/Fe])/3-[Mg/Fe])$, as a function of [Mg/Fe]. Our sample stars are color-coded by  metallicity (see color bar at right). The sample of metal-poor halo stars from JINAbase \citep{jinabase2018}, which includes \citet{roederer2014b}, \citet{rpa3}, and \citet{yong2013}, are shown with gray dots. The theoretical yields of CCSNe and HNe are from \citet{nomoto2013}.}
\label{crmndalpha}
\end{figure*}

Iron-peak elements at low metallicities are primarily produced through nucleosynthesis processes involving both complete and incomplete Si burning in CCSNe. The left panel in Figure~\ref{crmndalpha} shows the distribution of [Mn/Fe] versus [Cr/Fe], which are formed via incomplete Si burning in CCSNe. Following \citet{heger2002,heger2010}, \citet{qian2002}, \citet{nomoto2013}, and \citet{kobayashi2020}, very massive stars ($80 < M/{M}_{\odot} < 240$) belonging to Population III explode as pair-instability supernovae (PISNe), which should not produce a correlation between [Mn/Fe] and [Cr/Fe]. However, our results show a correlation between these two ratios, albeit with variability among stars from the literature and our current sample. The GC escapee marked in green stands out as a clear outlier to the trend. The Pearson correlation coefficient between [Cr/Fe] and [Mn/Fe] is 0.407, indicating a moderate, positive correlation. The presence of this correlation suggests that PISNe are unlikely to be the dominant progenitors of these stars. Therefore, based on this relationship, CCSNe associated with moderately high-mass stars ($M/{M}_{\odot} < 80$) are likely the primary contributors to the interstellar medium during the formation epoch of these stars. Below, we further investigate the role of CCSNe using the $\alpha$-elements.

As detailed in Section\,\ref{sec:alpha-elem}, we present findings on several stars exhibiting low Mg levels alongside either normal or enhanced abundances of other $\alpha$-elements such as Ca, Si, and Ti. At low metallicities, $\alpha$-elements are produced during hydrostatic burning and explosive nucleosynthesis phase of CCSNe. While O and Mg arise during hydrostatic burning in massive stars ($M/M_{\odot} > 35$), Ca, Si, and Ti originate during explosive burning in slightly less-massive stars ($M/M_{\odot} < 25$) \citep{kobayashi2020,mucci2023}. Thus, we divide the $\alpha$-elements into the two groups.

The right panel in Figure \ref{crmndalpha} illustrates the distribution of our sample stars, color-coded by [Fe/H], alongside the extensively studied samples from \citet{yong2013} and \citet{roederer2014b} in the \\${\rm \Delta\alpha = [(Ca/Fe+Si/Fe+Ti/Fe)/3-Mg/Fe}]$ vs. [Mg/Fe] plane. Notably, the stars demonstrate a discernible trend, with $\Delta\alpha$ decreasing as [Mg/Fe] increases which is also noted for the data from literature. The upward trend observed in $\Delta\alpha$ towards lower [Mg/Fe] could signify a likely increasing contribution from Type Ia supernovae at higher metallicities as evidenced by the prevalence of metal-rich stars shown in blue towards the upper left region. However, it is important to consider that the decreasing trend observed at lower overall $\alpha$-element abundances may not be solely attributed to an increase in SN Ia material. There is a possibility that higher mass CCSNe also contribute to the gas from which these stars form, especially in dwarf galaxy environments characterized by lower star formation rates and bottom heavy initial mass function \citep{mcwilliam2018}. Conversely, stars with higher [Mg/Fe] ratios ($> +0.4$) suggest increased contributions from massive core-collapse supernovae (CCSNe), with potential contributions from hypernovae (HNe) \citep{kobayashi2020}. For reference, we also show predicted yields for CCSNe progenitors of 15 and 25\,M$_{\odot}$ for $Z=0$, based on the yields provided by \citet{heger2010}, and \citet{limongi} as discussed by \citet{nomoto2013}. The [Mg/Fe] values tend to increase while $\Delta\alpha$ tends to decrease with increasing mass for CCSNe models. However, the integrated stellar yields across the IMF for a given metallicity are crucial in determining the chemical evolution of the galaxy. Hence, we have also incorporated models of Type II supernova with contribution from hypernovae taken from \citet{nomoto2013}, weighted by the Salpeter IMF for a mass range of 0.07--50 M\(\odot\), for different metallicities, represented by color-coded filled squares. As seen in the figure, they are found to have lower $\Delta\alpha$ with relatively higher [Mg/Fe]. Consequently, we infer that the stars manifesting larger disparities between the two groups of $\alpha$-element could be more likely to be born from gas enriched by massive CCSNe and HNe (for $\Delta\alpha < -0.2$) or Type Ia supernovae (for $\Delta\alpha > +0.4$).

\begin{figure*}
\centering
\includegraphics[width=2.0\columnwidth]{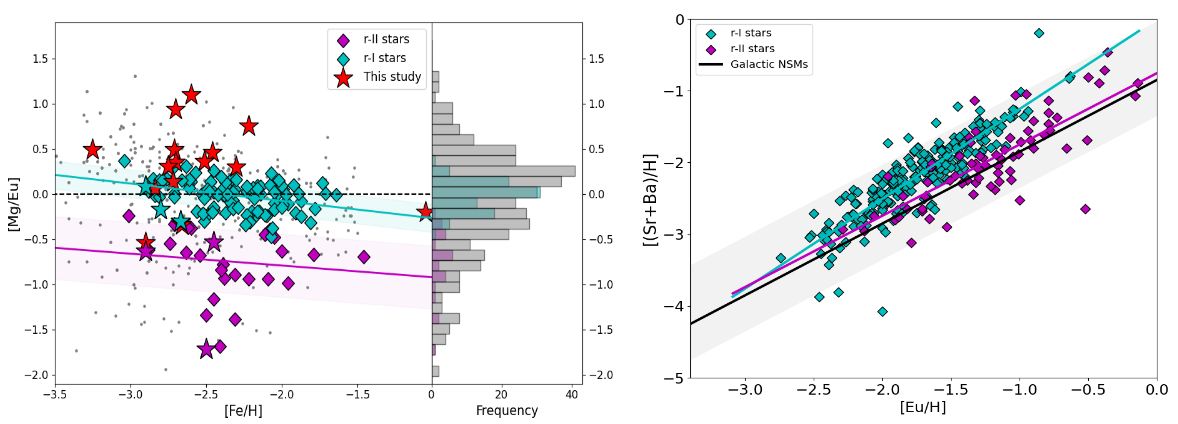}
\caption{Left panel: Distribution of [Mg/Eu] as a function of [Fe/H] for $r$-I, $r$-II and non-RPE stars. The red stars are the non-RPE stars from this study, the $r$-I and $r$-II stars are taken from \citet{rpa1}, \citet{rpa2}, \citet{rpa3} and \citet{rpa4} and the gray-filled circles are the stars from \citet{jinabase2018}. The $r$-I and $r$-II stars from this study are marked in the corresponding color. The histogram at the right compares the distributions of [Mg/Eu] from these sources over all metallicities. 
Right panel: The positive correlation of [(Sr+Ba)/H], as a function of [Eu/H]. The black solid line indicate sthe yields for main $r$-process from NSMs.}
\label{rproc}
\end{figure*}

\subsection{Sites and Evolution of the $r$-process}\label{rpsite}
 
The variation of [Mg/Eu] with metallicity can indicate the relative evolution of the two different astrophysical processes with time \citep{Naiman2018,banchemo} as they are dependent on the environment and can also be used to identify accretion history \citep{monty2024}. The left panel in Figure \ref{rproc} shows the distribution of [Mg/Eu], as a function of [Fe/H], for  different categories of stars ($r$-I, $r$-II, and non-RPE), accompanied by a histogram showing the distribution of [Mg/Eu] for these stars. The non-RPE halo stars from the literature are shown by gray dots, while the $r$-I and $r$-II stars from \citet{rpa2} and \citet{rpa3} are shown in cyan-filled and magenta-filled diamonds, respectively. The stars from this study are shown with red-filled stars. 

We can see differences between the $r$-I and $r$-II stars in the figure. The $r$-I stars exhibit a tight correlation around [Mg/Eu] = 0 over the given metallicity range, whereas the scatter is visibly much larger for $r$-II stars. A large part of the scatter comes from the stars having different birth environments, with different chemical enrichment histories. 
The larger scatter in [Mg/Eu] for the $r$-II stars can also attributed to the initial definition of $r$-II stars (defined by [Eu/Fe] $ > +0.7$; \citealt{rpa4}), while (by definition) the $r$-I stars only cover a 0.4 dex range in [Eu/Fe]. 

A decrease in the [Mg/Eu] ratios for both subsets of stars with increasing [Fe/H] is noted, and shown by the linear regression fits in the left panel of Figure \ref{rproc}. As discussed in \cite{Yutaka1}, such trends at low metallicity can provide valuable constraints on the timescale of $r$-process enrichment. The distribution of the non-RPE halo stars (gray circles) peak at [Mg/Eu] = +0.3, as seen in the histogram, while the the $r$-I stars peak at [Mg/Eu] = 0.0, and the  $r$-II stars peak at [Mg/Eu] $= -0.65$, which is expected based on their definitions. The large scatter and trends could be indicative of multiple production sites and regimes (e.g., the high and low [Mg/Eu] peaks as seen in the associated histogram) for these elements in the early Galaxy, which requires further investigations with simulations and chemical-evolution models. Hence, the $r$-process production for the $r$-II stars, at least as deduced from this work, as well as from other RPA studies, appears to be distinct from the non RPE counterparts; the origin of Eu in $r$-II stars is unlikely to be CCSNe, as signified by low [Mg/Eu]. More data are required, in particular for the $r$-II stars at [Fe/H] $\leq$ $-$3.0, to draw a more firm conclusion.

In order to probe deeper into the origin of the different neutron-capture elements discussed above, the right panel in Figure \ref{rproc} plots the distribution of [(Sr+Ba)/H], as a function of [Eu/H]\footnote{Here, \[\text{[(Sr+Ba)/H]} = \log \left( 10^{\log \epsilon (\text{Sr})} + 10^{\log \epsilon (\text{Ba})} \right) - \text{[(Sr+Ba)/H]}_{\odot},\] \[\text{[(Sr+Ba)/H]}_{\odot} = \log \left( 10^{\log \epsilon (\text{Sr}_{\odot})} + 10^{\log \epsilon (\text{Ba}_{\odot})} \right)\].}, thus removing the dependence on the metallicity. Any correlation (or lack thereof) between [(Sr+Ba)/H] vs. [Eu/H] is useful to derive constraints on the origins of these elements, as they are potentially produced in different astrophysical sites under different conditions \citep{tsuji1,Placco20rave,Bandyo2020,mardinimp23}. From inspection , the combined abundances of Sr and Ba correlate positively with Eu. However, the $r$-I and $r$-II stars exhibit a slight offset, which increases as a function of [Eu/H]. The yields for the main $r$-process due to Galactic NSMs is taken from \citet{erikansm},  shown with a black-solid line, which is very similar to the trend we find $r$-II stars to have. However, most of the $r$-I and $r$-II stars lie within the one sigma of the median yields region shaded in gray. The separation between the $r$-I and $r$-II stars sets in around [Eu/H] = $-3.0$ and increases with increasing [Eu/H], as indicated by the regression fits to the data. This could suggest an actual nucleosynthesis difference between the two sub-populations instead of dilution with metallicity.

The higher [(Sr+Ba)/H]  abundances in $r$-I stars, compared with those of the $r$-II stars, may also indicate different enrichment histories for these subsets of RPE stars.
However, we also note that the correlation between [(Sr+Ba)/H] vs. [Eu/H] is not necessarily produced with common astrophysical sites, as they could also be produced continuously from different sites in the Milky Way as discussed in \cite{Yutaka2}. The deviation of the trends for $r$-I stars from Galactic NSMs might indicate other sources of $r$-process enrichment in the early Galaxy.

\subsection{Likely Globular Cluster Escapees in the sample}

\begin{figure*}
\includegraphics[width=1.01\columnwidth]{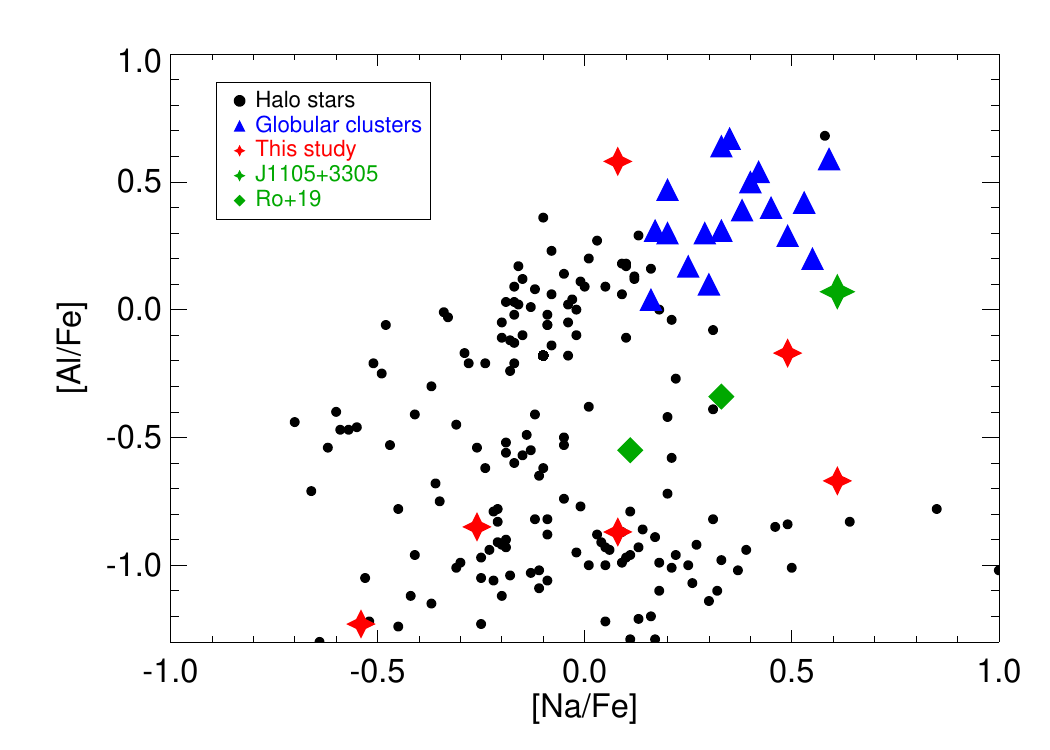}
\includegraphics[width=1.01\columnwidth]{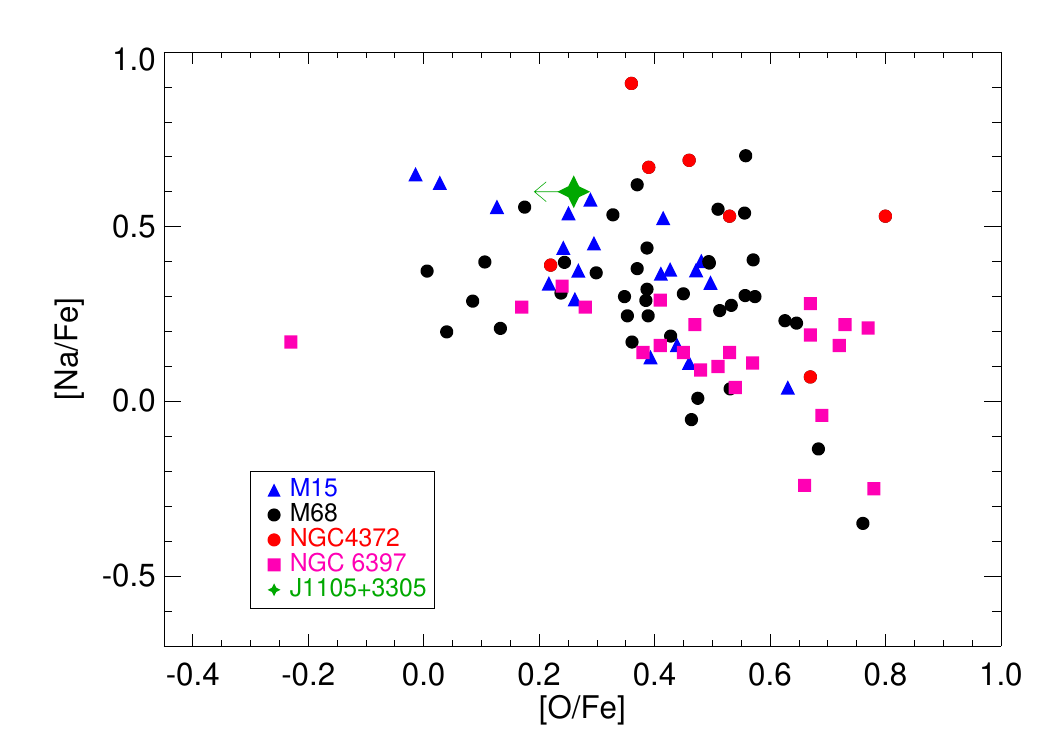}
\caption{Left panel: Distribution of the LTE abundances for Al and Na for stars in the halo and GCs. The black dots mark the halo stars obtained from \citet{sudasaga}, including  
\citet{carretta2009} and \citet{bangc}, the blue triangles denote the GC stars from \citet{carretta2009}. The stars from this study are shown in red; the likely GC escapee 2MASS J11052721+3305150 is marked with a green-filled star while other potential GC escapees from \citet{Roe2019gc} are marked in green-filled diamonds.
Right panel: The position of 2MASS J11052721+3305150 is compared to bonafide GC stars from several clusters in the anti-correlated Na-O plane; based on the O upper limit, it is seen to fall along the GC population.}
\label{gce}
\end{figure*}

The GCs and the halo field populations exhibit similar trends in the abundances of $\alpha$-, Fe-peak, and neutron-capture elements \citep{gratton2004,pritzl2005,gratton2012,lind2015,bangc}. However, many stars in GCs exhibit certain unique abundance ratios for the light elements, which are usually not found in halo stars (\citealt{krafta,norris79,dantona2019}, and numerous studies since). These chemical anomalies are thought to emerge as a result of self-pollution within the GCs (see \citealt{blardo2018} for a review). The light elements (Na, Mg, and Al, along with C) could be measured for the majority of the stars in this study. We find a handful of stars with peculiarities in their individual light-element abundances. For classification as a potential GC escapee based on the chemical abundances, stars are expected to show signatures of  elevated Na, Al, N, and (slightly) depleted Mg, C, and O abundances (see Figure \ref{gce}). Based on abundances alone, stars with such signatures are likely to be GC escapees, i.e., stars that were born in a GC but may have escaped the tidal radius either due to evaporation  or dissolution of their parent cluster over dynamical timescales. Such objects have been discovered in the halo in a number of studies \citep{martell2010,carollo2013,lind2015,martellapogee2016,Schiavon2017,rpa2,bangc,fern2021}.

One of our stars, 2MASS J11052721+3305150, with a metallicity of [Fe/H] $= -3.00$, strongly exhibits these anti-correlations, with LTE abundances of [Al/Fe] $ = +0.10$, [Na/Fe] $ = +0.55$, [Mg/Fe] $ = +0.32$, and [C/Fe] $= -0.21$. The LTE abundances of this star are shown in Figure \ref{gce}, along with those of the halo and GC populations from \citet{sudasaga} and \citet{carretta2009}, as described in \citet{bangc}. C abundances for this star are corrected for stellar evolutionary effects based on the calculations by \citet{placco2014}. This star falls distinctly closer to the GC population, despite its low Fe contentcompared to typical GCs. Such anomalies in the lighter elements could be attributed to signatures of advanced hydrogen burning in the star-forming clouds that result from the mixing of the ejecta of a progenitor population with the undiluted ISM. We also note that there are two other stars closer to the GC population, based on their Na and Al abundances, but their higher [Mg/Fe] and [C/Fe] does not allow us to associate them uniquely with GCs. 2MASS J11052721+3305150 could be among the most metal-poor GC escapees that have been reported to date. Stars of GC origin but with [Fe/H] much lower than the metallicity floor of GCs have also been reported in studies by \citet{Roe2019gc}, \citet{c19}, and \cite{sestito}. As shown in the left panel of Figure \ref{crmndalpha}, this star also does not follow the Mn-Cr correlation; it may also have received contributions from more massive supernovae. It is not possible to dynamically constrain the host GC for this star, as GCs at such low metallicites might no longer exist in the Galaxy. We also note the derived upper limit of [Eu/Fe] $<$ $-$0.27 is lower than that of GC stars, which may also indicate a lack of $r$-process enrichment at such low metallicities in GCs. We also note that determination of O abundances could conclusively associate the abundances of such objects with second-generation GC stars. Similar discoveries in the future will place better constraints on the metallicity floor and formation timescales of Galactic GCs.

\subsection{CEMP Stars in the Sample}

Of the five CEMP stars in our sample, four are CEMP-no (carbon-enhanced stars without enhancements of neutron-capture elements), and one is a CEMP-$s$ star (carbon-enhanced stars with enhancements of $s$-process elements). At very and extremely low metallicites, the CEMP-no stars dominate the halo population \citep{carollo2014,Yoon2016,yoon2018,lee2019}, and hence are extremely important for understanding the nature of their progenitors and early supernovae \citep{Ban2018,Sku2024}. The origin of carbon in CEMP-no stars has so far been shown to be to be intrinsic to the birth cloud of the stars, that is, not associated with mass transfer from a binary companion \citep{star14,Hansen2016a,Hansen2016b}. 

\begin{figure*}
\centering
\includegraphics[width=1.99\columnwidth,scale=0.5]{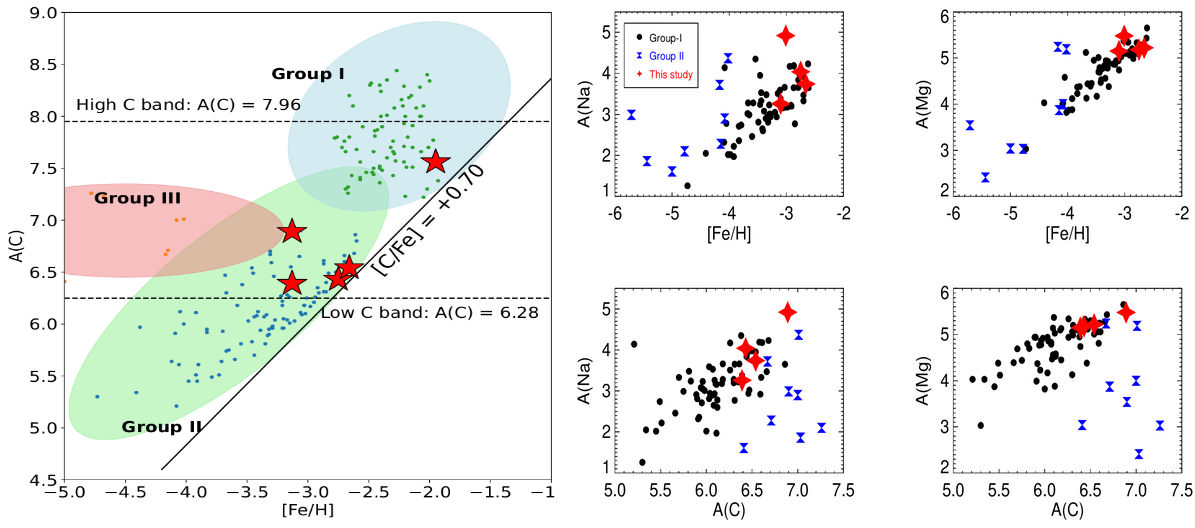}
\caption{Left panel: Classification of CEMP stars. The CEMP-no stars fall among the Group II and Group III stars in the low-C band while the CEMP-$s$ stars fall among the Group I stars in the high-C band. The new CEMP stars from this study are shown in red. Right panel: 
Distribution of the light element Na and the $\alpha$-element Mg for the Group II (black symbols) and Group III (blue symbols) CEMP-no stars, as a function of [Fe/H] (upper right panel) and $A$(C) (lower right panel). The literature data are taken from \citet{Yoon2016}.}
\label{cemp}
\end{figure*}

The Yoon-Beers Diagram \citep{Yoon2016} shown in the left panel of Figure \ref{cemp} indicates that the CEMP-no stars occupy the regions marked as Group II and Group III, while the Group I stars are predominantly of the CEMP-$s$ type. The stars in this study are marked in red. The CEMP-no stars fall closer to the lower C-band, and three of them are bonafide Group II stars, which are commonly associated with mixing and fallback supernovae progenitors \citep{nomoto2013,Maeder2015,Yoon2016}. However, one star (2MASS J22175058+2104371) can be associated with either the Group II or Group III populations based on its position in the figure. Stars belonging to Group III are rare, and 2MASS J22175058+2104371 presents an opportunity for a dedicated study to derive the abundances for other key elements including N and O, to understand the elevated C levels and uncover more details about its nucleosynthesis history and the formation channels. It would also be important to probe into the binary nature of the CEMP stars \citep{agbsus} with multi-epoch observations.

The right panel of Figure \ref{cemp} shows the distribution of the absolute abundances of Na and Mg for the Group II and Group III stars, marked with black and blue, respectively, as functions of [Fe/H] and $A$(C). The Na and Mg abundances in Group II stars scale with both metallicity and carbon abundances, whereas the Group III stars exhibit no clear trends. Three stars in this study marked in red fall among the Group II population, while 2MASS J22175058+2104371 occupies a distinct position, and could not be associated with certainty into one of the groups.

\section{Summary and Conclusions}\label{sec:conc}

This study focuses on faint (down to $V = 15.8$) VMP/EMP stars (down to [Fe/H] = $-3.30$) stars from the $R$-Process Alliance, observed with the HORuS spectrograph on the Gran Telescopio Canarias. Detailed chemical abundances of light, $\alpha$-, Fe-peak, and neutron-capture elements, along with upper limits whenever feasible, for 41 stars are reported, shedding light on the origin, evolution, and chemical-enrichment history of these stars.

This paper highlights the discovery of 1 limited-$r$, 3 $r$-I, 4 $r$-II star, and 5 CEMP stars, each with implications for subsequent dedicated follow-up studies with higher resolution and SNR spectra. The identification of a star possibly of a globular cluster origin at an extremely low metallicity ([Fe/H] = -3.0) could invigorate a re-evaluation of the metallicity floor for GCs. We also report the discovery of 6 new Mg-poor stars, and results for 23 new VMP/EMP stars.

The elemental abundance ratios of our sample of stars are compared to literature values, including the results from homogeneous studies by \cite{yong2013}, \cite{roederer2014b}, \cite{rpa1}, \cite{rpa2}, \citet{rpa3}, and \citet{rpa4}. Despite a larger scatter, similar trends for various abundance ratios with metallicity were found. 
Using the presence of a moderate correlation between Cr and Mn, we demonstrate a lack of significant contribution from PISNe to the early star-forming gas. Instead,  this indicates a prevalence of slightly less-massive CCSNe as dominant contributors of elements in the very early Universe.

In Section \ref{alphaa}, we presented findings on stars with low Mg levels alongside Solar-level or enhanced abundances of other $\alpha$-elements such as Ca, Si, and Ti. These groups elements are primarily produced by CCSNe during hydrostatic (Mg, O) and explosive nucleosynthesis (Ca, Si and Ti). We categorize the $\alpha$-elements into these two groups accordingly to investigate the discrepancy.Our analysis suggests that stars showing significant differences in $\alpha$-element abundances may originate from gas enriched by massive core-collapse supernovae and hypernovae (for $\Delta\alpha < -0.2$) or from Type Ia supernovae (for $\Delta\alpha > +0.4$).

Our exploration of the sites and evolution of $r$-process elements in section \ref{rpsite} reveals distinct trends in [Mg/Eu] relative to [Fe/H]. Both stellar sub-populations exhibit a decreasing trend in [Mg/Eu] as [Fe/H] increases. While $r$-I stars show a tight correlation around [Mg/Eu] = 0, $r$-II stars exhibit a larger scatter and a more pronounced decrease with increasing [Fe/H]. These findings suggest multiple production regimes of these elements in the early Galaxy, indicating a decoupling of $r$-process production from CCSNe, particularly for $r$-II stars. We also find a positive correlation between [(Sr+Ba)]/H and [Eu/H], which suggests a shared site for the enrichment of $r$-process elements in the early Galaxy, despite with a slight offset between the trends for $r$-I and $r$-II stars. The yields for Galactic NSMs are found to be similar to the $r$-II stars and are slightly deviated for the $r$-I stars. 

This paper identifies a star potentially escaping from a globular cluster, based on its light-element abundance ratios, offering insights into the metallicity thresholds and formation timescales of Galactic GCs. This paper also indentifies five new CEMP stars, which are important for understanding the nature of the progenitor population and early supernovae.

Upcoming work from the RPA, based on a homogeneous analysis of  $\sim 2000$ RPE and non-RPE stars, will provide further statistical evidence that should verify our current results, as well as place constraints on the origin(s) of RPE stars by probing the contributions from various nucleosynthesis channels in the early Galaxy.

\acknowledgements


We thank an anonymous referee for thoughtful suggestions that helped us to improve our paper and tighten its presentation.
C.A.P. is thankful for financial support from the Spanish Ministry MICINN projects AYA2017-86389-P and PID2020-117493GB-I00. T.C.B. acknowledges support from grant PHY 14-30152; Physics Frontier Center/JINA Center for the Evolution of the Elements (JINA-CEE), and from OISE-1927130: The International Research  Network for Nuclear Astrophysics (IReNA), awarded by the US National Science Foundation. A.F. acknowledge support from NSF grant AST-2307436.
T.T.H. acknowledges support from the Swedish Research Council (VR 2021-05556). E.M.H acknowledges this work performed under the auspices of the U.S. Department of Energy by Lawrence Livermore National Laboratory under Contract DE-AC52-07NA27344. Document release number: LLNL-JRNL-2000129. The work of V.M.P. is supported by NOIRLab, which is managed by the Association of Universities for Research in Astronomy (AURA) under a cooperative agreement with the National Science Foundation. I.U.R. acknowledges support from the U.S. National Science Foundation (grants PHY~14-30152, Physics Frontier Center/JINA-CEE; AST~1815403; and AST~2205847 I.U.R.), and the NASA Astrophysics Data Analysis Program (grant 80NSSC21K0627).


\bibliography{main}{}
\bibliographystyle{aasjournal}

\startlongtable
\begin{deluxetable*}{l r l r r r c r c}
\tablewidth{100pt}
\tabletypesize{\footnotesize}
\tablecaption{\label{tab:linelist} Atomic Line Properties, Equivalent Widths, 
Absolute Abundances (before corrections), and Measurement Uncertainties of the Target Stars}
\tablehead{
\colhead{Star ID} & \colhead{Species} & \colhead{$\lambda$} & \colhead{$\chi$} & \colhead{$\log\,gf$} & \colhead{EW} & \colhead{EW error} & \colhead{$A(X)$}\\
& & \colhead{({\AA})}      & \colhead{(eV)} &     &\colhead{(m{\AA})} &\colhead{(m{\AA})}   &  } 
\startdata
2MASS J22424551+2720245 &Na I &5889.950 &0.000 &0.110 &67.01 &0.40 &2.452\\
2MASS J22424551+2720245 &Na I &5895.920 &0.000 &$-$0.190 &50.84 &0.35 &2.452\\
2MASS J22424551+2720245 &Mg I &4167.270 &4.350 &$-$0.740 &14.32 &0.21 &4.627\\
2MASS J22424551+2720245 &Mg I &4702.990 &4.330 &$-$0.440 &28.11 &0.42 &4.650\\
2MASS J22424551+2720245 &Mg I &5528.400 &4.350 &$-$0.550 &18.05 &0.34 &4.509\\
2MASS J22424551+2720245 &Al I &3961.520 &0.010 &$-$0.330 &57.53 &0.45 &1.967\\
2MASS J22424551+2720245 &Ca I &4283.010 &1.890 &$-$0.200 &25.47 &0.50 &3.457\\
2MASS J22424551+2720245 &Ca I &4434.960 &1.890 &$-$0.060 &27.07 &0.37 &3.345\\
2MASS J22424551+2720245 &Ca I &4435.690 &1.890 &$-$0.550 &5.65 &0.18 &3.005\\
2MASS J22424551+2720245 &Ca I &4454.780 &1.900 &0.260 &31.93 &0.30 &3.148\\
2MASS J22424551+2720245 &Ca I &5588.760 &2.520 &0.300 &16.64 &0.48 &3.342\\
2MASS J22424551+2720245 &Ca I &5601.290 &2.530 &$-$0.570 &2.12 &0.12 &3.236\\
2MASS J22424551+2720245 &Ca I &6122.220 &1.890 &$-$0.330 &20.01 &0.73 &3.351\\
2MASS J22424551+2720245 &Ca I &6162.170 &1.900 &$-$0.110 &32.62 &0.32 &3.453\\
2MASS J22424551+2720245 &Ca I &6439.070 &2.520 &0.330 &17.76 &0.48 &3.322\\
2MASS J22424551+2720245 &Sc II &4246.820 &0.320 &0.240 &77.72 &0.31 &0.248\\
2MASS J22424551+2720245 &Sc II &4320.730 &0.600 &$-$0.250 &33.88 &0.44 &$-$0.178\\
2MASS J22424551+2720245 &Sc II &4324.980 &0.590 &$-$0.440 &47.90 &0.30 &0.322\\
2MASS J22424551+2720245 &Sc II &4400.390 &0.600 &$-$0.540 &49.13 &0.52 &0.452\\
2MASS J22424551+2720245 &Sc II &4670.410 &1.360 &$-$0.580 &4.76 &0.13 &$-$0.074\\
2MASS J22424551+2720245 &Sc II &5031.010 &1.360 &$-$0.400 &16.40 &0.35 &0.328\\
2MASS J22424551+2720245 &Sc II &5526.790 &1.770 &0.020 &8.85 &0.31 &0.031\\
2MASS J22424551+2720245 &Ti I &4512.730 &0.840 &$-$0.400 &1.55 &0.15 &1.663\\
2MASS J22424551+2720245 &Ti I &4548.760 &0.830 &$-$0.280 &7.07 &0.15 &2.219\\
2MASS J22424551+2720245 &Ti I &4681.910 &0.050 &$-$1.010 &8.70 &0.24 &2.151\\
2MASS J22424551+2720245 &Ti I &4981.730 &0.840 &0.570 &28.81 &0.55 &2.120\\
2MASS J22424551+2720245 &Ti I &4991.070 &0.840 &0.450 &13.21 &0.37 &1.781\\
2MASS J22424551+2720245 &Ti II &4337.910 &1.080 &$-$0.960 &50.82 &0.30 &1.878\\
2MASS J22424551+2720245 &Ti II &4399.770 &1.240 &$-$1.200 &37.10 &0.28 &1.960\\
2MASS J22424551+2720245 &Ti II &4417.710 &1.170 &$-$1.190 &28.99 &0.22 &1.680\\
2MASS J22424551+2720245 &Ti II &4444.550 &1.120 &$-$2.200 &8.98 &0.18 &1.971\\
2MASS J22424551+2720245 &Ti II &4450.480 &1.080 &$-$1.520 &22.96 &0.35 &1.751\\
2MASS J22424551+2720245 &Ti II &4501.270 &1.120 &$-$0.770 &58.79 &0.29 &1.926\\
2MASS J22424551+2720245 &Ti II &4571.970 &1.570 &$-$0.310 &53.76 &0.31 &1.835\\
2MASS J22424551+2720245 &Ti II &4708.660 &1.240 &$-$2.350 &3.61 &0.17 &1.811\\
2MASS J22424551+2720245 &Ti II &5129.160 &1.890 &$-$1.340 &4.08 &0.16 &1.569\\
2MASS J22424551+2720245 &Ti II &5188.690 &1.580 &$-$1.050 &18.96 &0.19 &1.683\\
2MASS J22424551+2720245 &V I &4389.980 &0.280 &0.220 &15.38 &0.22 &1.491\\
2MASS J22424551+2720245 &V II &3997.110 &1.480 &$-$1.200 &12.33 &0.16 &1.481\\
2MASS J22424551+2720245 &V II &4023.380 &1.800 &$-$0.610 &19.71 &0.17 &1.511\\
2MASS J22424551+2720245 &Cr I &4274.800 &0.000 &$-$0.220 &49.17 &0.21 &1.709\\
2MASS J22424551+2720245 &Cr I &520C40 &0.940 &0.020 &25.91 &0.31 &1.889\\
2MASS J22424551+2720245 &Cr I &5298.280 &0.980 &$-$1.140 &7.53 &0.27 &2.417\\
2MASS J22424551+2720245 &Cr I &5345.800 &1.000 &$-$0.950 &7.59 &0.28 &2.252\\
2MASS J22424551+2720245 &Cr II &4588.200 &4.070 &$-$0.650 &4.47 &0.20 &2.503\\
2MASS J22424551+2720245 &Mn I &4030.750 &0.000 &$-$0.500 &3C0 &0.26 &0.990\\
2MASS J22424551+2720245 &Mn I &403Li60 &0.000 &$-$0.650 &40.95 &0.26 &1.262\\
2MASS J22424551+2720245 &Mn I &4034.480 &0.000 &$-$0.840 &38.42 &0.42 &1.388\\
2MASS J22424551+2720245 &Fe I &4132.060 &1.610 &$-$0.680 &67.70 &0.32 &4.142\\
2MASS J22424551+2720245 &Fe I &4143.870 &1.560 &$-$0.510 &82.33 &0.27 &4.398\\
2MASS J22424551+2720245 &Fe I &4181.760 &2.830 &$-$0.370 &22.61 &0.18 &3.923\\
2MASS J22424551+2720245 &Fe I &4187.040 &2.450 &$-$0.560 &39.55 &0.27 &4.110\\
2MASS J22424551+2720245 &Fe I &4187.800 &2.420 &$-$0.510 &35.70 &0.27 &3.933\\
2MASS J22424551+2720245 &Fe I &4191.430 &2.470 &$-$0.670 &28.48 &0.32 &3.970\\
2MASS J22424551+2720245 &Fe I &4216.180 &0.000 &$-$3.360 &40.93 &0.25 &4.151\\
2MASS J22424551+2720245 &Fe I &4238.810 &3.400 &$-$0.230 &23.13 &0.57 &4.426\\
2MASS J22424551+2720245 &Fe I &4250.120 &2.470 &$-$0.380 &41.46 &0.30 &3.993\\
2MASS J22424551+2720245 &Fe I &4250.790 &1.560 &$-$0.710 &75.47 &0.28 &4.353\\
2MASS J22424551+2720245 &Fe I &4260.470 &2.400 &0.080 &68.83 &0.30 &4.215\\
2MASS J22424551+2720245 &Fe I &4271.150 &2.450 &$-$0.340 &41.63 &0.27 &3.932\\
2MASS J22424551+2720245 &Fe I &4282.400 &2.180 &$-$0.780 &40.96 &0.39 &4.063\\
2MASS J22424551+2720245 &Fe I &4352.730 &2.220 &$-$1.290 &19.63 &0.32 &4.055\\
2MASS J22424551+2720245 &Fe I &4415.120 &1.610 &$-$0.620 &81.94 &0.31 &4.473\\
2MASS J22424551+2720245 &Fe I &4427.310 &0.050 &$-$2.920 &68.85 &0.28 &4.592\\
2MASS J22424551+2720245 &Fe I &4430.610 &2.220 &$-$1.730 &21.92 &0.60 &4.553\\
2MASS J22424551+2720245 &Fe I &4442.340 &2.200 &$-$1.230 &28.73 &0.29 &4.208\\
2MASS J22424551+2720245 &Fe I &4447.720 &2.220 &$-$1.360 &16.17 &0.47 &4.004\\
2MASS J22424551+2720245 &Fe I &4461.650 &0.090 &$-$3.190 &60.69 &0.35 &4.609\\
2MASS J22424551+2720245 &Fe I &4466.550 &2.830 &$-$0.600 &20.99 &0.29 &4.082\\
2MASS J22424551+2720245 &Fe I &4489.740 &0.120 &$-$3.900 &24.04 &0.25 &4.377\\
2MASS J22424551+2720245 &Fe I &4494.560 &2.200 &$-$1.140 &26.07 &0.30 &4.046\\
2MASS J22424551+2720245 &Fe I &4531.150 &1.480 &$-$2.100 &22.62 &0.34 &4.094\\
2MASS J22424551+2720245 &Fe I &4592.650 &1.560 &$-$2.460 &16.33 &0.29 &4.350\\
2MASS J22424551+2720245 &Fe I &4602.940 &1.490 &$-$2.210 &26.03 &0.54 &4.302\\
2MASS J22424551+2720245 &Fe I &4733.590 &1.490 &$-$2.990 &4.61 &0.17 &4.160\\
2MASS J22424551+2720245 &Fe I &4736.770 &3.210 &$-$0.670 &7.44 &0.20 &4.010\\
2MASS J22424551+2720245 &Fe I &4871.320 &2.870 &$-$0.340 &36.12 &0.37 &4.219\\
2MASS J22424551+2720245 &Fe I &4872.140 &2.880 &$-$0.570 &27.81 &0.40 &4.260\\
2MASS J22424551+2720245 &Fe I &4882.140 &3.420 &$-$1.480 &4.50 &0.19 &4.809\\
2MASS J22424551+2720245 &Fe I &4890.760 &2.880 &$-$0.380 &21.89 &0.36 &3.914\\
2MASS J22424551+2720245 &Fe I &4891.490 &2.850 &$-$0.110 &38.02 &0.26 &4.009\\
2MASS J22424551+2720245 &Fe I &4910.020 &3.400 &$-$1.280 &3.75 &0.13 &4.502\\
2MASS J22424551+2720245 &Fe I &4918.990 &2.860 &$-$0.340 &26.40 &0.23 &3.969\\
2MASS J22424551+2720245 &Fe I &4920.500 &2.830 &0.070 &51.78 &0.36 &4.138\\
2MASS J22424551+2720245 &Fe I &4938.810 &2.880 &$-$1.080 &9.99 &0.37 &4.182\\
2MASS J22424551+2720245 &Fe I &4939.690 &0.860 &$-$3.250 &21.77 &0.62 &4.477\\
2MASS J22424551+2720245 &Fe I &4946.390 &3.370 &$-$1.110 &2.98 &0.24 &4.191\\
2MASS J22424551+2720245 &Fe I &5001.860 &3.880 &$-$0.010 &8.26 &0.16 &4.132\\
2MASS J22424551+2720245 &Fe I &5005.710 &3.880 &$-$0.120 &18.38 &0.41 &4.665\\
2MASS J22424551+2720245 &Fe I &5006.120 &2.830 &$-$0.620 &19.32 &0.32 &4.015\\
2MASS J22424551+2720245 &Fe I &5051.630 &0.920 &$-$2.760 &31.13 &0.34 &4.285\\
2MASS J22424551+2720245 &Fe I &5123.720 &1.010 &$-$3.060 &13.41 &0.26 &4.172\\
2MASS J22424551+2720245 &Fe I &5127.360 &0.920 &$-$3.250 &21.15 &0.67 &4.515\\
2MASS J22424551+2720245 &Fe I &5131.470 &2.220 &$-$2.520 &4.11 &0.14 &4.441\\
2MASS J22424551+2720245 &Fe I &5150.840 &0.990 &$-$3.040 &15.21 &0.30 &4.196\\
2MASS J22424551+2720245 &Fe I &5151.910 &1.010 &$-$3.320 &7.88 &0.28 &4.164\\
2MASS J22424551+2720245 &Fe I &5166.280 &0.000 &$-$4.120 &21.03 &0.47 &4.315\\
2MASS J22424551+2720245 &Fe I &5202.340 &2.180 &$-$1.870 &7.09 &0.24 &3.998\\
2MASS J22424551+2720245 &Fe I &5215.180 &3.270 &$-$0.860 &14.37 &0.75 &4.573\\
2MASS J22424551+2720245 &Fe I &5216.270 &1.610 &$-$2.080 &20.28 &0.38 &4.105\\
2MASS J22424551+2720245 &Fe I &5217.390 &3.210 &$-$1.070 &8.88 &0.20 &4.470\\
2MASS J22424551+2720245 &Fe I &5232.940 &2.940 &$-$0.060 &32.69 &0.38 &3.915\\
2MASS J22424551+2720245 &Fe I &5266.560 &3.000 &$-$0.380 &18.17 &0.34 &3.916\\
2MASS J22424551+2720245 &Fe I &5281.790 &3.040 &$-$0.830 &10.45 &0.26 &4.116\\
2MASS J22424551+2720245 &Fe I &5283.620 &3.240 &$-$0.450 &14.66 &0.28 &4.138\\
2MASS J22424551+2720245 &Fe I &5341.020 &1.610 &$-$1.950 &25.10 &0.37 &4.099\\
2MASS J22424551+2720245 &Fe I &5371.490 &0.960 &$-$1.640 &90.69 &0.33 &4.869\\
2MASS J22424551+2720245 &Fe I &5397.130 &0.920 &$-$1.980 &69.21 &0.41 &4.486\\
2MASS J22424551+2720245 &Fe I &5405.770 &0.990 &$-$1.850 &79.74 &0.37 &4.771\\
2MASS J22424551+2720245 &Fe I &5415.200 &4.390 &0.640 &14.07 &0.54 &4.303\\
2MASS J22424551+2720245 &Fe I &5497.520 &1.010 &$-$2.820 &24.65 &0.35 &4.258\\
2MASS J22424551+2720245 &Fe I &5501.470 &0.960 &$-$3.050 &24.65 &0.43 &4.430\\
2MASS J22424551+2720245 &Fe I &5506.780 &0.990 &$-$2.790 &29.47 &0.32 &4.324\\
2MASS J22424551+2720245 &Fe I &5586.760 &3.370 &$-$0.110 &21.58 &1.21 &4.150\\
2MASS J22424551+2720245 &Fe I &5662.520 &4.180 &$-$0.410 &7.40 &0.26 &4.786\\
2MASS J22424551+2720245 &Fe II &4178.860 &2.580 &$-$2.510 &16.20 &0.25 &4.182\\
2MASS J22424551+2720245 &Fe II &4515.340 &2.840 &$-$2.600 &10.21 &0.14 &4.294\\
2MASS J22424551+2720245 &Fe II &4555.890 &2.830 &$-$2.400 &19.87 &0.42 &4.442\\
2MASS J22424551+2720245 &Fe II &4583.830 &2.810 &$-$1.940 &34.39 &0.49 &4.333\\
2MASS J22424551+2720245 &Fe II &4620.520 &2.830 &$-$3.210 &2.02 &0.14 &4.130\\
2MASS J22424551+2720245 &Fe II &5234.630 &3.220 &$-$2.180 &7.00 &0.20 &4.076\\
2MASS J22424551+2720245 &Fe II &5276.000 &3.200 &$-$2.010 &11.41 &0.40 &4.122\\
2MASS J22424551+2720245 &Co I &4118.770 &1.050 &$-$0.480 &42.41 &0.59 &2.394\\
2MASS J22424551+2720245 &Co I &4121.320 &0.920 &$-$0.330 &24.83 &0.25 &1.643\\
2MASS J22424551+2720245 &Ni I &4604.990 &3.480 &$-$0.240 &5.51 &0.19 &3.440\\
2MASS J22424551+2720245 &Ni I &4714.420 &3.380 &0.250 &12.55 &0.27 &3.239\\
2MASS J22424551+2720245 &Ni I &5081.110 &3.850 &0.300 &10.10 &0.34 &3.581\\
2MASS J22424551+2720245 &Ni I &5084.080 &3.680 &0.030 &7.29 &0.15 &3.501\\
2MASS J22424551+2720245 &Ni I &5476.900 &1.830 &$-$0.780 &21.27 &0.60 &2.772\\
2MASS J22424551+2720245 &Zn I &4810.540 &4.080 &$-$0.150 &2.83 &0.24 &1.311\\
\enddata
\tablecomments{The linelist for one star is shown here. The full table for all the targets will be made available electronically.}
\end{deluxetable*}
\begin{longrotatetable}
\begin{deluxetable*}{l r r r r r r r r r r}
\tablewidth{0pt}
\tabletypesize{\footnotesize}
\tablecaption{\label{tab:abund_light}Light Element Abundances}
\tablehead{
\\
\colhead{Star ID}  & \colhead{$\mathrm{[Na I/H]}$} & \colhead{$\mathrm{[Mg I/H]}$} & \colhead{$\mathrm{[Al I/H]}$}& \colhead{$\mathrm{[Si I/H]}$} & \colhead{$\mathrm{[Ca I/H]}$} & \colhead{$\mathrm{[Sc II/H]}$} & \colhead{$\mathrm{[Ti I/H]}$} & \colhead{$\mathrm{[Ti II/H]}$}
}
\startdata
2MASS J00125284+4726278 & $-2.12\pm0.18$ & $-2.14\pm0.33$ & $-3.81\pm0.25$ & \dots & $ -2.05\pm	0.29 $ & $-2.43\pm0.23$ & $-2.39\pm0.22$ & $-2.30\pm0.26$ \\    
2MASS J01171437+2911580 & $-1.96\pm0.16$ & $-2.07\pm0.21$ & $-3.98\pm0.15$ & $-1.77\pm0.20$ & $ -2.07\pm	0.19 $ & $-2.49\pm0.23$ & $-2.39\pm0.17$ & $-2.17\pm0.19$ \\    
2MASS J01261714+2620558 & $-1.36\pm0.18$ & $-0.99\pm0.20$ & $-2.33\pm0.20$ & $-0.10\pm0.47$ & $ -0.55\pm	0.21 $ & $ 0.21\pm0.23$ & $-0.61\pm0.35$ & $-0.12\pm0.44$ \\   
2MASS J02462013$-$1518418 & $-2.55\pm0.25$ & $-2.44\pm0.15$ & $-4.13\pm0.35$ & $0.19\pm0.26$ & $ -2.56\pm	0.14 $ & $-2.92\pm0.18$ & $-2.63\pm0.16$ & $-2.64\pm0.28$ \\    
2MASS J04051243+2141326 & $-2.47\pm0.14$ & $-2.27\pm0.12$ & $3.78\pm0.25$ & \dots & $ -1.92\pm	0.35 $ & $-2.64\pm0.28$ & $-2.07\pm0.12$ & $-2.44\pm0.28$ \\    
2MASS J04464970+2124561 & $-1.67\pm0.11$ & $-2.04\pm0.20$ & \dots & $-1.40\pm0.15$ & $ -1.30\pm	0.29 $ & $-1.35\pm0.17$ & $-1.43\pm0.27$ & $-1.26\pm0.32$ \\    
2MASS J05455436+4420133 & $-2.43\pm0.13$ & $-2.41\pm0.25$ & \dots & \dots & $ -2.53\pm	0.19 $ & $-2.94\pm0.34$ & $-2.62\pm0.18$ & $-2.46\pm0.35$ \\    
2MASS J06114434+1151292 & $-2.43\pm0.14$ & $-2.60\pm0.44$ & \dots & $-1.26\pm0.28$ & $ -2.27\pm	0.17 $ & $-2.89\pm0.14$ & $-2.64\pm0.16$ & $-2.67\pm0.32$ \\    
2MASS J06321853+3547202 & $-2.16\pm0.10$ & $-2.45\pm0.12$ & $-4.33\pm0.17$ & $-1.68\pm0.12$ & $ -2.43\pm	0.15 $ & $-2.75\pm0.17$ & $-2.36\pm0.24$ & $-2.36\pm0.29$ \\    
2MASS J07424682+3533180 & $-2.03\pm0.17$ & $-2.38\pm0.14$ & $-4.04\pm0.15$ & $-1.35\pm0.22$ & $ -2.19\pm	0.29 $ & $-2.59\pm0.30$ & $-2.40\pm0.16$ & $-2.29\pm0.12$ \\    
2MASS J07532819+2350207 & $-3.00\pm0.10$ & $-2.70\pm0.20$ & $-3.59\pm0.15$ & $-1.52\pm0.22$ & $ -2.21\pm	0.25 $ & $-2.21\pm0.28$ & $-2.01\pm0.28$ & $-2.62\pm0.36$ \\    
2MASS J08011752+4530033 & $-2.48\pm0.12$ & $-2.44\pm0.19$ & $-4.34\pm0.18$ & $-2.36\pm0.30$ & $ -2.32\pm	0.24 $ & $-2.97\pm0.18$ & $-2.51\pm0.21$ & $-2.50\pm0.18$ \\    
2MASS J08203890+3619470 & $-2.03\pm0.13$ & $-2.30\pm0.15$ & \dots & \dots & $ -2.39\pm	0.44 $ & $-2.15\pm0.20$ & $-2.34\pm0.19$ & $-2.35\pm0.21$ \\    
2MASS J08471988+3209297 & $-1.37\pm0.15$ & $-2.16\pm0.22$ & \dots & $-1.39\pm0.19$ & $ -1.80\pm	0.23 $ & $-2.07\pm0.16$ & $-1.79\pm0.35$ & $-1.51\pm0.24$ \\    
2MASS J09092839+1704521 & $-1.79\pm0.13$ & $-1.89\pm0.27$ & $-4.74\pm0.25$ & $-1.65\pm0.16$ & $ -1.66\pm	0.25 $ & $-1.89\pm0.34$ & $-2.07\pm0.29$ & $-1.91\pm0.23$ \\    
2MASS J09143307+2351544 & $-2.76\pm0.11$ & $-2.82\pm0.30$ & $-3.42\pm0.14$ & $-2.81\pm0.60$ & $ -2.78\pm	0.27 $ & $-3.31\pm0.14$ & $-3.09\pm0.40$ & $-2.81\pm0.29$ \\    
2MASS J09185208+5107215 & $-2.98\pm0.08$ & $-2.45\pm0.32$ & $-2.48\pm0.15$ & $-2.12\pm0.10$ & $ -2.41\pm	0.27 $ & $-3.13\pm0.15$ & $-2.68\pm0.30$ & $-2.94\pm0.24$ \\    
2MASS J09261148+1802142 & $-2.38\pm0.14$ & $-2.35\pm0.12$ & \dots & $-1.48\pm0.16$ & $ -2.25\pm	0.27 $ & \dots & $-2.46\pm0.30$ & $-2.25\pm0.30$ \\    
2MASS J09563630+5953170 & $-1.69\pm0.13$ & $-1.99\pm0.14$ & \dots & $-1.52\pm0.10$ & $ -1.90\pm	0.29 $ & $-1.84\pm0.15$ & $-1.53\pm0.28$ & $-1.70\pm0.25$ \\    
2MASS J10122279+2716094 & \dots & $-1.99\pm0.18$ & $-2.02\pm0.10$ & $-1.43\pm0.10$ & $ -1.66\pm	0.27 $ & $-2.22\pm0.17$ & $-2.21\pm0.13$ & $-2.23\pm0.22$ \\    
2MASS J10542923+2056561 & $-0.76\pm0.12$ & $-0.89\pm0.13$ & \dots & $-0.32\pm0.16$ & $ -0.23\pm	0.18 $ & $ 0.31\pm0.13$ & $-0.07\pm0.24$ & $ 0.20\pm0.18$ \\  
2MASS J11052721+3305150 & $-2.40\pm0.40$ & $-2.63\pm0.14$ & $-2.90\pm0.15$ & \dots & $ -2.46\pm	0.23 $ & $-2.56\pm0.17$ & $-2.17\pm0.17$ & $-2.20\pm0.24$ \\    
2MASS J12131230+2506598 & $-1.83\pm0.17$ & $-2.38\pm0.18$ & \dots & \dots & $ -2.05\pm	0.50 $ & $-1.93\pm0.28$ & $-2.51\pm0.17$ & $-2.54\pm0.19$ \\    
2MASS J12334194+1952177 & $-2.77\pm0.18$ & $-2.26\pm0.11$ & \dots & $-1.78\pm0.16$ & $ -2.67\pm	0.33 $ & $-2.52\pm0.16$ & $-2.75\pm0.18$ & $-2.34\pm0.29$ \\    
2MASS J12445815+5820391 & $-2.68\pm0.23$ & $-2.38\pm0.22$ & $-3.63\pm0.22$ & $-2.46\pm0.15$ & $ -2.29\pm	0.20 $ & $-2.48\pm0.21$ & $-2.86\pm0.20$ & $-2.41\pm0.20$ \\    
2MASS J13281307+5503080 & $0.22\pm0.17$ & $-0.17\pm0.25$ & \dots & $0.52\pm0.34$ & $ 0.26\pm	 0.23 $ & $ 0.85\pm0.26$ & $ 0.15\pm0.22$ & $ 0.31\pm0.17$ \\
2MASS J13525684+2243314 & $-3.32\pm0.43$ & $-2.12\pm0.14$ & $-4.95\pm0.16$ & $-1.22\pm0.15$ & $ -2.01\pm	0.25 $ & $-1.96\pm0.17$ & $-2.32\pm0.33$ & $-2.17\pm0.34$ \\    
2MASS J13545109+3820077 & $-2.01\pm0.10$ & $-2.10\pm0.21$ & $-4.05\pm0.15$ & \dots & $ -2.07\pm	0.25 $ & $-2.73\pm0.21$ & $-2.33\pm0.25$ & $-2.49\pm0.32$ \\    
2MASS J14245543+2707241 & $-0.48\pm0.11$ & $-1.96\pm0.21$ & \dots & $ 0.15\pm0.21$ & $ -1.74\pm	0.30 $ & $-1.15\pm0.33$ & $-1.09\pm0.27$ & $-1.40\pm0.26$ \\   
2MASS J14445238+4038527 & $-2.44\pm0.26$ & $-2.09\pm0.18$ & \dots & $-1.35\pm0.20$ & $ -1.98\pm	0.36 $ & $-1.91\pm0.20$ & $-2.06\pm0.21$ & $-1.92\pm0.40$ \\    
2MASS J15442141+5735135 & $-1.76\pm0.32$ & $-2.36\pm0.22$ & $-5.19\pm0.25$ & \dots & $ -2.31\pm	0.38 $ & $-3.00\pm0.34$ & $-2.62\pm0.34$ & $-2.59\pm0.41$ \\    
2MASS J16374570+3230413 & $-1.83\pm0.16$ & $-2.16\pm0.13$ & $-3.11\pm0.18$ & $-1.28\pm0.10$ & $ -2.05\pm	0.29 $ & $-2.24\pm0.29$ & $-2.09\pm0.29$ & $-2.22\pm0.21$ \\    
2MASS J16380702+4059136 & $-1.94\pm0.14$ & $-2.58\pm0.25$ & $-4.44\pm0.19$ & \dots & $ -2.00\pm	0.26 $ & $-2.37\pm0.15$ & $-2.10\pm0.26$ & $-2.09\pm0.40$ \\    
2MASS J16393877+3616077 & $-1.26\pm0.14$ & $-2.17\pm0.18$ & \dots & $-1.24\pm0.14$ & $ -1.54\pm	0.33 $ & $-1.45\pm0.41$ & $-1.58\pm0.20$ & $-1.41\pm0.30$ \\    
2MASS J16451495+4357120 & $-2.20\pm0.11$ & $-2.42\pm0.20$ & \dots & $-2.30\pm0.15$ & $ -2.50\pm	0.25 $ & $-3.49\pm0.14$ & $-2.66\pm0.24$ & $-2.82\pm0.27$ \\    
2MASS J17041197+1626552 & $-2.50\pm0.24$ & $-2.38\pm0.16$ & $-4.49\pm0.25$ & $-1.82\pm0.30$ & $ -1.99\pm	0.30 $ & $-2.31\pm0.13$ & $-2.16\pm0.25$ & $-2.05\pm0.20$ \\    
2MASS J17045729+3720576 & $-2.50\pm0.51$ & $-2.16\pm0.09$ & \dots & $-1.70\pm0.21$ & $ -2.34\pm	0.23 $ & $-2.01\pm0.02$ & $-2.07\pm0.25$ & $-1.74\pm0.31$ \\    
2MASS J17125701+4432051 & $-2.45\pm0.13$ & $-2.32\pm0.20$ & \dots & $-1.77\pm0.13$ & $ -2.29\pm	0.24 $ & $-2.62\pm0.15$ & $-2.41\pm0.16$ & $-2.48\pm0.25$ \\    
2MASS J21463220+2456393 & $-0.97\pm0.24$ & $-0.94\pm0.04$ & \dots & \dots & $ -0.47\pm	0.34 $ & $-0.67\pm0.18$ & $-0.29\pm0.42$ & $-0.56\pm0.33$ \\    
2MASS J22175058+2104371 & $-1.32\pm0.24$ & $-2.12\pm0.07$ & \dots & \dots & $ -2.37\pm	0.32 $ & $-2.89\pm0.32$ & $-2.13\pm0.37$ & $-2.70\pm0.37$ \\    
2MASS J22424551+2720245 & $-3.79\pm0.15$ & $-3.00\pm0.16$ & $-4.48\pm0.15$ & \dots & $ -3.04\pm	0.14 $ & $-2.99\pm0.22$ & $-2.96\pm0.22$ & $-3.14\pm0.13$ \\  
\enddata
\end{deluxetable*}
\end{longrotatetable}

\begin{longrotatetable}
\begin{deluxetable*}{l r r r r r r r r r r}
\tablewidth{0pt}
\tabletypesize{\footnotesize}
\tablecaption{\label{tab:abund_fe_peak}Fe-Peak Element Abundances}
\tablehead{
\\
\colhead{Star ID} & \colhead{$\mathrm{[V I/H]}$} & \colhead{$\mathrm{[V II/H]}$} & \colhead{$\mathrm{[Cr I/H]}$} & \colhead{$\mathrm{[Cr II/H]}$} & \colhead{$\mathrm{[Mn I/H]}$}& \colhead{$\mathrm{[Co I/H]}$} & \colhead{$\mathrm{[Ni I/H]}$} & \colhead{$\mathrm{[Cu/H]}$} & \colhead{$\mathrm{[Zn I/H]}$}
}
\startdata
2MASS J00125284+4726278 & $-1.75\pm 0.11$ & \dots & $-2.73\pm0.23$ & $-2.68\pm0.16$ & $-2.92\pm0.28$ & & $-2.38\pm0.25$ & \dots & $-2.46\pm0.17$\\       
2MASS J01171437+2911580 & $-2.97\pm 0.10$ & \dots & $-2.66\pm0.22$ & $-2.52\pm0.15$ & $-2.97\pm0.32$ & $ -2.53\pm0.21 $ & $-2.64\pm0.16$ & \dots & $-2.53\pm0.14$\\    
2MASS J01261714+2620558 & $-0.51\pm 0.15$ & $-0.31\pm0.12$ & $-0.62\pm0.38$ & $-0.31\pm0.50$ & $-0.80\pm0.36$ & $ -0.66\pm	0.52 $ & $-1.07\pm0.41$ & $-0.24\pm0.10$ & $-0.46\pm0.16$\\    
2MASS J02462013$-$1518418 & \dots & $-2.22\pm0.15$ & $-3.30\pm0.28$ & \dots & $-3.35\pm0.32$ & $ -3.21\pm0.13 $ & $-2.76\pm0.17$ & \dots & $-2.65\pm0.16$\\     
2MASS J04051243+2141326 & $-1.35\pm 0.10$ & \dots & $-3.14\pm0.17$ & $-2.18\pm0.26$ & $-3.13\pm0.25$ & $ -2.45\pm0.14 $ & $-2.01\pm0.17$ & \dots & $-2.14\pm0.18$ \\   
2MASS J04464970+2124561 & \dots & \dots & $-2.14\pm0.28$ & $-1.82\pm0.17$ & $-1.65\pm0.12$ & $ -1.92\pm	0.14 $ & $-1.67\pm0.24$ & \dots & $-1.19\pm0.15$\\     
2MASS J05455436+4420133 & $-2.78\pm 0.18$ & $-2.61\pm0.23$ & $-2.89\pm0.24$ & $-2.32\pm0.15$ & $-3.18\pm0.18$ & \dots & $-2.75\pm0.22$ & \dots & \dots  \\      
2MASS J06114434+1151292 & \dots & \dots & $-3.00\pm0.19$ & $-2.63\pm0.14$ & $-3.51\pm0.17$ & $ -3.10\pm0.28 $ & $-2.50\pm0.21$ & \dots & $-2.50\pm0.13$ \\    
2MASS J06321853+3547202 & $-2.74\pm 0.12$ & \dots & $-2.84\pm0.17$ & \dots & $-3.53\pm0.18$ & $ -2.15\pm	0.14 $ & $-2.86\pm0.18$ & \dots & \dots \\    
2MASS J07424682+3533180 & $-2.35\pm 0.10$ & \dots & $-2.65\pm0.43$ & $-1.91\pm0.30$ & $-2.80\pm0.42$ & $ -1.80\pm	0.15 $ & $-2.60\pm0.13$ & \dots & $-2.57\pm0.14$\\    
2MASS J07532819+2350207 & $-0.95\pm 0.14$ & $-1.61\pm0.11$ & $-2.89\pm0.30$ & \dots & $-2.61\pm0.45$ & $ -2.28\pm	0.12 $ & $-1.88\pm0.14$ & \dots & $-1.99\pm0.11$\\    
2MASS J08011752+4530033 & $-2.48\pm 0.10$ & $-2.64\pm0.19$ & $-3.09\pm0.19$ & \dots & $-3.28\pm0.19$ & $ -2.45\pm	0.22 $ & $-2.70\pm0.18$ & \dots & $-2.64\pm0.30$\\    
2MASS J08203890+3619470 & $-3.19\pm 0.10$ & \dots & $-3.06\pm0.32$ & \dots & $-3.38\pm0.10$ & $ -2.62\pm	0.10 $ & $-2.42\pm0.20$ & $-2.94\pm0.10$ & $-1.41\pm0.10$\\    
2MASS J08471988+3209297 & \dots & \dots & $-2.20\pm0.17$ & $-1.86\pm0.17$ & $-2.43\pm0.37$ & \dots & $-2.27\pm0.34$ & \dots & \dots \\       
2MASS J09092839+1704521 & $-2.12\pm 0.14$ & \dots & $-2.44\pm0.19$ & \dots & $-3.04\pm0.42$ & $ -2.13\pm	0.36 $ & $-2.21\pm0.11$ & \dots & $-2.55\pm0.12$\\    
2MASS J09143307+2351544 & $-3.16\pm 0.10$ & \dots & $-3.52\pm0.30$ & $-2.66\pm0.20$ & $-3.82\pm0.42$ & $ -3.48\pm	0.21 $ & $-2.95\pm0.16$ & \dots & \dots \\    
2MASS J09185208+5107215 & $-2.91\pm 0.11$ & $-2.63\pm0.36$ & $-2.79\pm0.34$ & \dots & $-3.66\pm0.30$ & $ -1.95\pm	0.37 $ & $-2.67\pm0.30$ & $-2.72\pm0.10$ & $-2.50\pm0.23$\\    
2MASS J09261148+1802142 & $-3.20\pm 0.15$ & \dots & $-2.78\pm0.13$ & $-2.50\pm0.36$ & $-2.27\pm0.21$ & $ -1.63\pm	0.15 $ & $-2.57\pm0.18$ & \dots & $-2.47\pm0.10$\\    
2MASS J09563630+5953170 & $-2.36\pm 0.19$ & $-2.20\pm0.15$ & $-1.96\pm0.15$ & $-2.08\pm0.25$ & $-2.44\pm0.32$ & $ -2.07\pm	1.04 $ & $-2.04\pm0.19$ & $-2.45\pm0.19$ & $-2.14\pm0.14$\\    
2MASS J10122279+2716094 & $-1.99\pm 0.10$ & \dots & $-2.44\pm0.30$ & $-2.43\pm0.20$ & $-3.05\pm0.53$ & $ -2.31\pm	0.18 $ & $-2.49\pm0.18$ & \dots & $2.65\pm0.10$\\    
2MASS J10542923+2056561 & $-0.70\pm 0.13$ & \dots & $-0.36\pm0.18$ & $-2.33\pm0.15$ & $-0.03\pm0.32$ & $ -0.43\pm	0.53 $ & $-0.53\pm0.12$ & $0.52\pm0.10$ & $-0.35\pm0.20$\\    
2MASS J11052721+3305150 & \dots & $-2.09\pm0.11$ & $-3.31\pm0.18$ & \dots & $-2.93\pm0.28$ & $ -2.62\pm	0.10 $ & $-2.41\pm0.20$ & \dots & $-1.98\pm0.14$\\     
2MASS J12131230+2506598 & $-2.51\pm 0.10$ & \dots & $-3.49\pm0.18$ & \dots & $-3.31\pm0.20$ & $ -1.02\pm	0.15 $ & $-2.55\pm0.27$ & \dots & $-2.46\pm0.16$\\    
2MASS J12334194+1952177 & $-2.70\pm 0.33$ & \dots & $-3.31\pm0.39$ & \dots & $-3.86\pm0.16$ & $ -3.01\pm	0.22 $ & $-2.84\pm0.25$ & \dots & $-2.38\pm0.29$\\    
2MASS J12445815+5820391 & \dots & \dots & $-3.04\pm0.26$ & $-2.08\pm0.19$ & $-2.74\pm0.18$ & \dots & $-2.60\pm0.16$ & \dots & $-2.31\pm0.12$\\       
2MASS J13281307+5503080 & $ 1.11\pm 0.10$ & \dots & $0.10\pm0.21$ & $ 0.21\pm0.49$ & $0.03\pm0.23$ & $ 0.48\pm	0.10 $ & $-0.10\pm0.21$ & $ 1.13\pm0.10$ & $ 0.29\pm0.17$\\
2MASS J13525684+2243314 & $-2.00\pm 0.15$ & \dots & $-2.59\pm0.30$ & $-2.02\pm0.20$ & $-3.56\pm0.22$ & $ -2.14\pm	0.62 $ & $-2.49\pm0.13$ & $-3.11\pm0.15$ & $-2.40\pm0.12$\\    
2MASS J13545109+3820077 & $-3.03\pm 0.12$ & \dots & $-2.79\pm0.35$ & $-2.51\pm0.21$ & $-3.02\pm0.17$ & $ -3.05\pm	0.10 $ & $-2.27\pm0.31$ & \dots & $-2.40\pm0.11$\\    
2MASS J14245543+2707241 & $-1.12\pm 0.10$ & \dots & $-1.16\pm0.14$ & $-1.25\pm0.13$ & $-1.65\pm0.28$ & $ -1.17\pm	0.14 $ & $-1.37\pm0.18$ & \dots & \dots \\    
2MASS J14445238+4038527 & \dots & \dots & $-2.51\pm0.25$ & \dots & & \dots & $-2.43\pm0.17$ & \dots & \dots \\        
2MASS J15442141+5735135 & $-2.56\pm 0.18$ & \dots & $-2.72\pm0.34$ & \dots & $-3.53\pm0.33$ & $ -2.42\pm	0.17 $ & $-2.72\pm0.17$ & \dots & $-2.18\pm0.21$\\    
2MASS J16374570+3230413 & \dots & $-2.46\pm0.13$ & $-2.77\pm0.28$ & $-2.41\pm0.10$ & $-2.83\pm0.13$ & $ -1.58\pm	0.14 $ & $-2.26\pm0.16$ & $-2.58\pm0.10$ & $-2.32\pm0.12$\\     
2MASS J16380702+4059136 & $-1.37\pm 0.15$ & $-1.43\pm0.18$ & $-2.56\pm0.14$ & $-2.09\pm0.13$ & $-2.45\pm0.10$ & $ -1.68\pm	0.15 $ & $-2.60\pm0.17$ & \dots & $-2.00\pm0.12$\\    
2MASS J16393877+3616077 & \dots & \dots & $-1.85\pm0.30$ & \dots & $-2.05\pm0.11$ & $ -2.34\pm	0.20 $ & $-2.11\pm0.14$ & \dots & $-1.65\pm0.27$\\     
2MASS J16451495+4357120 & \dots & \dots & $-2.58\pm0.22$ & $-2.28\pm0.11$ & $-2.66\pm0.36$ & $ -3.34\pm	0.12 $ & $-2.48\pm0.19$ & \dots & $-2.31\pm0.39$\\     
2MASS J17041197+1626552 & \dots & $-2.20\pm0.15$ & $-2.61\pm0.24$ & $-2.10\pm0.17$ & $-2.70\pm0.19$ & $ -3.06\pm	0.17 $ & $-2.32\pm0.32$ & $-2.43\pm0.14$ & $-2.29\pm0.11$\\     
2MASS J17045729+3720576 & \dots & \dots & $-2.69\pm0.36$ & \dots & $-2.13\pm0.16$ & $ -2.14\pm	0.17 $ & $-2.39\pm0.18$ & \dots & \dots \\     
2MASS J17125701+4432051 & $-1.87\pm 0.14$ & \dots & $-2.95\pm0.21$ & $-2.60\pm0.24$ & $-3.50\pm0.27$ & $ -3.16\pm	0.11 $ & $-2.44\pm0.22$ & \dots & $-2.41\pm0.14$\\    
2MASS J21463220+2456393 & $-1.21\pm 0.10$ & \dots & $-0.74\pm0.27$ & $-1.12\pm0.29$ & $-0.68\pm0.28$ & $ -1.05\pm	0.10 $ & $-1.03\pm0.27$ & \dots & $-1.12\pm0.12$ \\   
2MASS J22175058+2104371 & $-2.16\pm 0.10$ & $-2.12\pm0.32$ & $-3.27\pm0.17$ & $-2.88\pm0.10$ & $-2.96\pm0.30$ & $ -2.81\pm	0.08 $ & $-2.83\pm0.19$ & \dots & \dots \\    
2MASS J22424551+2720245 & $-2.44\pm 0.10$ & $-2.43\pm0.03$ & $-3.57\pm0.28$ & $-3.14\pm0.10$ & $-4.22\pm0.17$ & $ -2.97\pm	0.38 $ & $-2.91\pm0.29$ & \dots & $-3.25\pm0.14$ \\   
\enddata
\end{deluxetable*}
\end{longrotatetable}

\end{document}